\documentclass[fleqn,usenatbib]{mnras}

\usepackage{newtxtext,newtxmath}
\usepackage[T1]{fontenc}

\DeclareRobustCommand{\VAN}[3]{#2}
\let\VANthebibliography\thebibliography
\def\thebibliography{\DeclareRobustCommand{\VAN}[3]{##3}\VANthebibliography}

\usepackage{graphicx}	% Including figure files
\usepackage{amsmath}	% Advanced maths commands
\title[Engine-Fed kilonovae. I.]{
Engine-fed Kilonovae (Mergernovae) - \uppercase\expandafter{\romannumeral1}. 
Dynamical Evolution and Energy Injection / Heating Efficiencies}
\author[S. Ai, B. Zhang \& Z. Zhu]{
Shunke Ai$^{1,2}$\thanks{E-mail: ais1@unlv.nevada.edu},
Bing Zhang$^{1,2}$\thanks{E-mail: bing.zhang@unlv.edu}
and Zhaohuan Zhu$^{1,2}$\thanks{E-mail: zhaohuan.zhu@unlv.edu}
\\
$^{1}$Nevada Center for Astrophysics, University of Nevada Las Vegas, Las Vegas, NV 89154, USA\\
$^{2}$Department of Physics and Astronomy, University of Nevada Las Vegas, Las Vegas, NV 89154, USA\\
}

\date{Accepted XXX. Received YYY; in original form ZZZ}
\pubyear{2022}

\begin{document}
\label{firstpage}
\pagerange{\pageref{firstpage}--\pageref{lastpage}}
\maketitle

\begin{abstract}
A binary neutron star merger is expected to be associated by a kilonova, transient optical emission powered by radioactive decay of the neutron-rich ejecta. If the post-merger remnant is a long-lived neutron star, additional energy injection to the ejecta is possible. In this first paper of a series, we study the dynamical evolution of the engine-fed kilonova (mergernova) ejecta in detail. We perform a semi-analytical study of the problem by adopting a modified mechanical blastwave model that invokes interaction between a Poynting-flux-dominated flow and a non-magnetized massive ejecta. Shortly after the engine is turned on, a pair of shocks would be excited. The reverse shock quickly reaches the wind-acceleration region and disappears (in a few seconds), whereas the forward shock soon breaks out from the ejecta (in $10^2$ - $10^3$ seconds) and continues to propagate in the surrounding interstellar medium. Most of the energy injected into the blastwave from the engine is stored as magnetic energy and kinetic energy. The internal energy fraction is $f_{\rm int} < 0.3$ for an ejecta mass equal to $10^{-3}M_{\odot}$. Overall, the energy injecting efficiency $\xi$ is at most $\sim 0.6$ and can be as small as $\sim 0.04$ at later times. Contrary to the previous assumption, efficient heating only happens before the forward shock breaks out of the ejecta with a heating efficiency $\xi_t \sim (0.006 - 0.3)$, which rapidly drops to $\sim 0$ afterwards. The engine-fed kilonova lightcurves will be carefully studied in Paper II.
\end{abstract}

%%%%%%%%%%%%%%%%%%%%%%%%%%%%%%%%%%%%%%%%%%%%%%%%%%

%%%%%%%%%%%%%%%%% BODY OF PAPER %%%%%%%%%%%%%%%%%%
\section{Introduction}
Binary neutron star (BNS) mergers are believed to be accompanied with the ejection of neutron-rich materials, either due to tidal effect during the merger \citep{lattimer74,rosswog99,rosswog05} or due to outflow ejection from the accretion torus surrounding the post-merger central remnant \citep{metzger14}. This neutron-rich ejecta would power a UV-optical-IR transient due to $r$-process and radioactive decay of the heavy elements in the ejecta \citep{lipaczyski98}, which is termed as ``kilonova'' by \cite{metzger10} who found that the peak luminosity is $\sim 10^{41} {\rm erg/s}$, about a thousand times brighter than the Eddington luminosity of solar-mass objects. A kilonova candidate was first detected in the afterglow of the short gamma-ray burst (GRB) GRB 130603B \citep{tanvir13,berger13,fan13}, and a kilonova event was confirmed in the BNS merger gravitational wave event GW170817 \citep{GW170817-EM}.

The merger product of a BNS merger depends on the NS equation of state (EoS) and the total mass in the binary \citep{rosswog00,rezzolla10,lasky14,gao16,ai20}. If the NS EoS is stiff enough, a supramassive NS supported by its rigid rotation or even a stable NS could be produced \citep{cook94,lasota96,breu16,studzinska16,bozzola18}. Due to its rapid rotation, the NS likely carries a strong magnetic field so is likely a millisecond magnetar. Such a magnetar would spin down via magnetic dipole radiation and gravitational radiation \citep{shapiro83,zhang01} and release energy, either directly emitting photons via magnetic dissipation \citep{zhang13} or injects energy into the ejecta in the form of a Poynting flux, which both accelerates and heats up the ejecta \citep{gao13,yu13,metzgerpiro14}. The type of central remnant might also influence the ejecta mass and the r-process nucleosynthesis in the ejecta \citep{rosswog14,radice18}. 

\cite{yu13} first studied the dynamical evolution and radiation signature of the optical transient powered by a magnetar engine and found that it is typically brighter and peaks earlier than the standard kilonovae. They advocated a term "mergernova" to embrace both radio-activity-powered and engine powered transients. \cite{metzgerpiro14} also studied the magnetar-powered kilonova emission including the role of $e^\pm$ pairs on radiation. Later searches from the archival short GRB afterglow data revealed both traditional kilonovae (e.g. GRB 060614 \citep{yang15}, GRB 050709 \citep{jin16}) and more luminous events probably powered by magnetars (e.g. GRB 050724, GRB 061006, GRB 070714B, and GRB 080503 \citep{gao15,gao17a}). The kilonova AT 2017gfo associated with GW170817 demands both a blue component and a red component \citep[e.g.][]{arcavi17,chornock17,cowperthwaite17,drout17,evans17,gao17b,kasen17,kilpatrick17,nicholl17,shappee17,smartt17,tanvir17,villar17}. The early-peaked blue component with peak luminosity $\gtrsim 10^{42} {\rm erg \ s^{-1}}$, together with the late-peaked red component with peak luminoisty $\sim 10^{41}{\rm erg \ s^{-1}}$, indicate that the total mass of ejecta should be as high as $\sim 0.06M_{\rm \odot}$ \citep[e.g.][]{villar17},
which is greater than the typical mass expected from numerical simulations, e.g. $\sim 10^{-4} -  {\rm few} \times 10^{-2} M_{\odot}$ \citep{rezzolla11,rezzolla13,hotokezaka13,siegel17,fujibayashi18}. However, some simulations on multi-dimensional radiati transfer claim that the ejecta is not necessary to be that massive \citep{kawaguchi18,korobkin21}, which could be consistent with the numerical simulations on BNS merger. The possibility of a long-lived engine powering the kilonova  \citep{yu18,lisz18} and the afterglow \citep{piro19,troja20} has been discussed for this event, even though the underlying NS is constrained not to possess a strong magnetic field \citep{ai18}. 

In all the previous work studying engine-fed kilonovae (mergernovae) \citep{yu13,metzgerpiro14,wollaeger19}, the energy injection process is treated through parameterization, e.g. a free parameter $\xi$ is adopted to denote the fraction of the spindown energy that is transferred to the kilonova (mergernova) ejecta. This injected energy would be partially used to accelerate the ejecta and partially used to heat up the ejecta to increase the brightness of the kilonova (mergernova). As a result, there should be an additional efficiency parameter $\xi_t$ for ``thermalization''. In the approximate treatment of \cite{yu13}, these efficiency parameters were generally denoted as $\xi$, and a typical value $\xi = 0.3$ was adopted in the calculations. Physically, these parameters are dictated by the complicated dynamical evolution of system invoking the interaction between a Poynting-flux-dominated wind and a massive ejecta as well as a surrounding medium ahead of the shell, which has not been studied carefully in the literature.

In a series of two papers, we perform an in-depth study of the physics of engine-fed kilonovae (mergernovae). In this first paper, we will study the dynamical evolution and energy injection/heating efficiency of the process in great detail. The general physical picture of the energy injection process is described in Section \ref{sec:description}. A mechanical model, as well as the shock jump conditions, to treat the problem are introduced in Section \ref{sec:model}. We perform a MHD simulation with \emph{Athena++}, in Section \ref{sec:simulation}, to verify the physical picture of the problem and the validity of the mechanical model. Then, in Section \ref{sec:dynamics}, we apply the shock model to the problem with realistic parameters and calculate the dynamical evolution of the blastwave. The energy injection efficiency and heating efficiency for different blastwave evolution stages are studied in Section \ref{sec:efficiencies}. The results are summarized in Section \ref{sec:conclusions} with some discussion. Detailed calculations of the lightcurves of engine-fed kilonovae (mergernovae) will be presented in Paper II. 

\section{General picture} \label{sec:description}
The interaction between two fluids can be treated as a Riemann problem, in which shocks or rarefaction waves are produced depending on the velocities, densities and pressures of the two fluids \citep[e.g.][]{rezzolla13}. A contact discontinuity (CD) exists between the two fluids. In this paper, we ignore the r-process nucleosynthesis power and focus on the energy injection process from the central engine wind. The engine wind can drive a forward shock (FS) in the cold ejecta. The criteria for a reverse shock (RS) to  form in a magnetized wind is \citep{zhangkobayashi05}
\begin{eqnarray}
\sigma_w < \frac{8}{3} \Gamma_{\rm w,ej}^2 \frac{\rho_{\rm ej}}{\rho_w},
\label{eq:RScriteria}
\end{eqnarray}
where $\rho_{\rm ej}$, $\rho_w$, and $\Gamma_{\rm w,ej}$ represent density of the ejecta, density of the wind, and the relative bulk Lorentz factor between the wind and ejecta, respectively. The wind is assumed to be cold with a magnetization parameter $\sigma_w = B^2/(4\pi \rho c^2)$, where $B$ is the strength of magnetic field. All the quantities above, except the Lorentz factor, are defined in the rest frames of the respective fluids. Since the ejecta is extremely dense, a RS should still be formed even if $\sigma_w$ is high (see details in Section \ref{sec:region4}). When $\sigma_w$ is high enough, in the rest frame of the central engine, which is hereafter termed as the ``lab frame'', the RS would propagate towards the central engine. Usually, the blastwave is defined as the region between the RS and the FS. However, in this paper, since the RS may vanish, we define the region that have been historically shocked either by the FS or the RS as the blastwave.

The schematic picture for the energy injecting process is shown in Figure \ref{fig:schematic}. The whole process can be divided into three stages:

\begin{figure*}
\begin{tabular}{ll}
\resizebox{95mm}{!}{\includegraphics[]{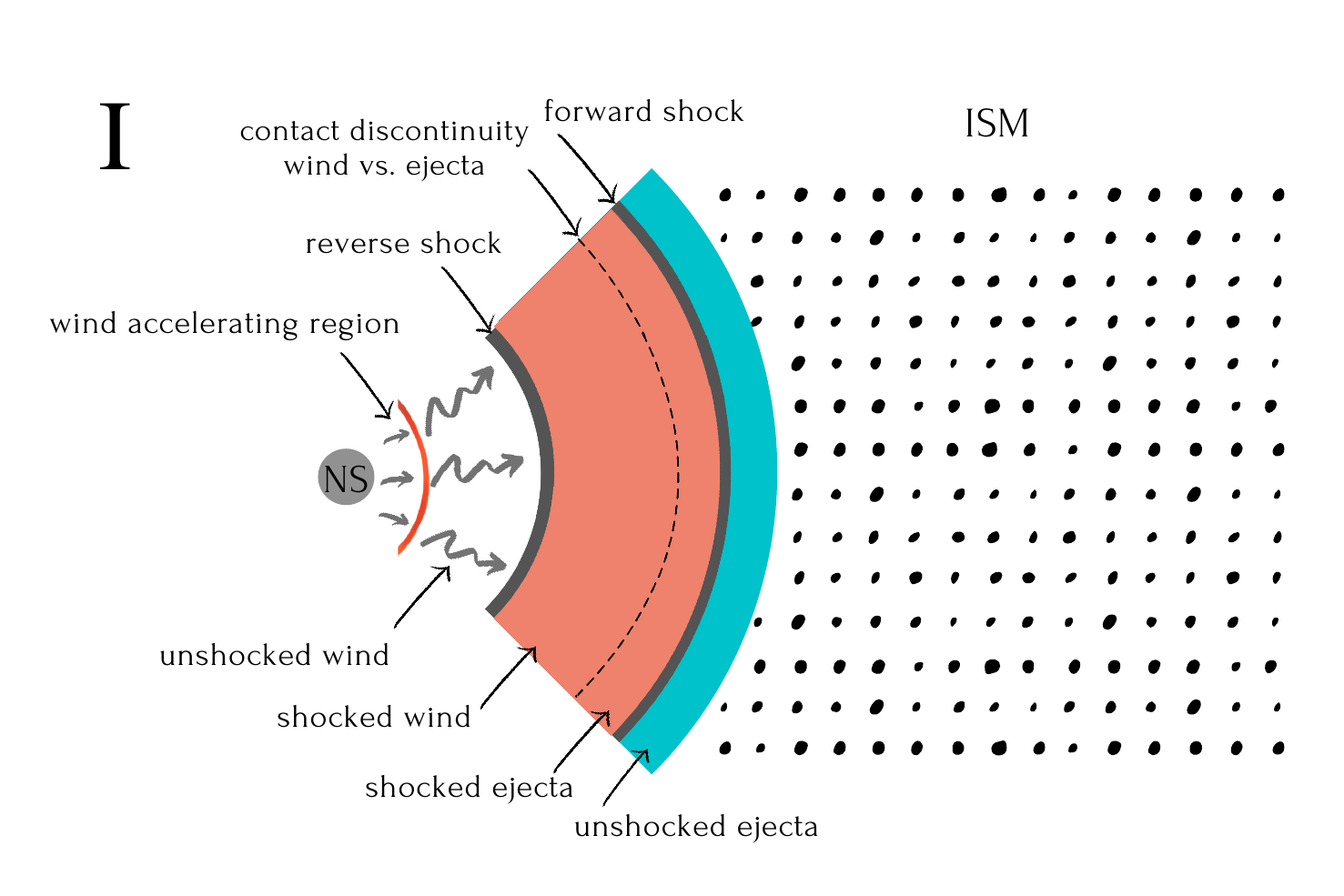}} &
\resizebox{90mm}{!}{\includegraphics[]{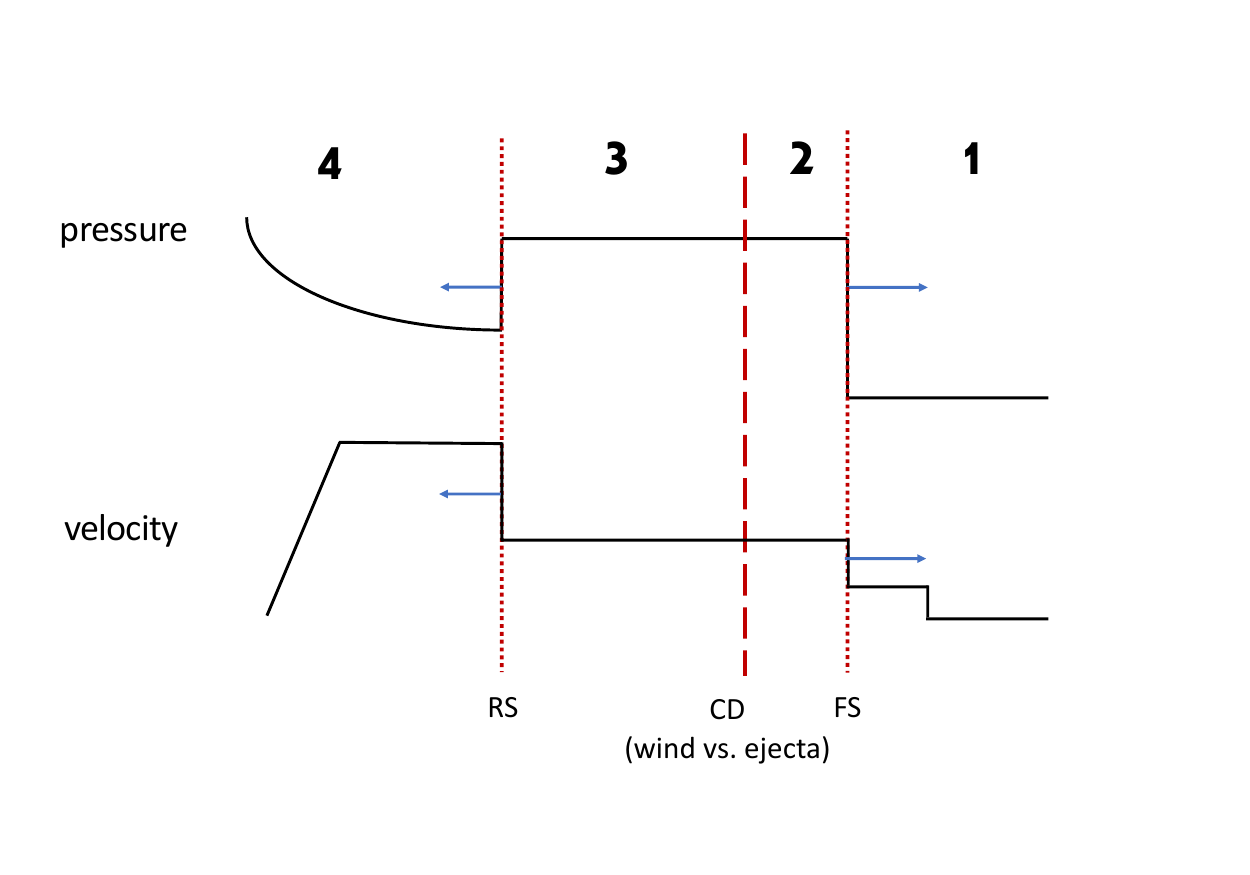}} \\
\resizebox{95mm}{!}{\includegraphics[]{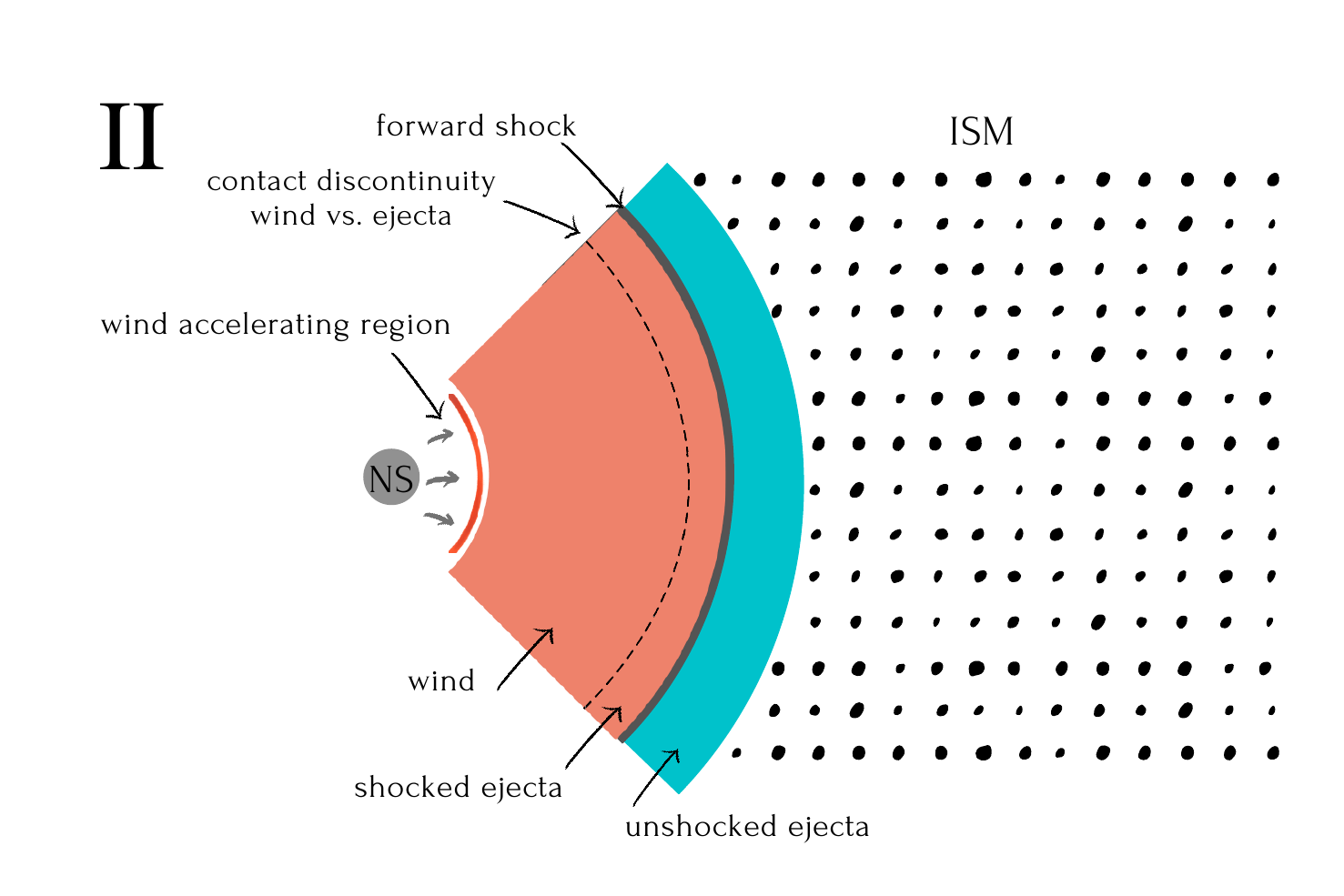}} &
\resizebox{90mm}{!}{\includegraphics[]{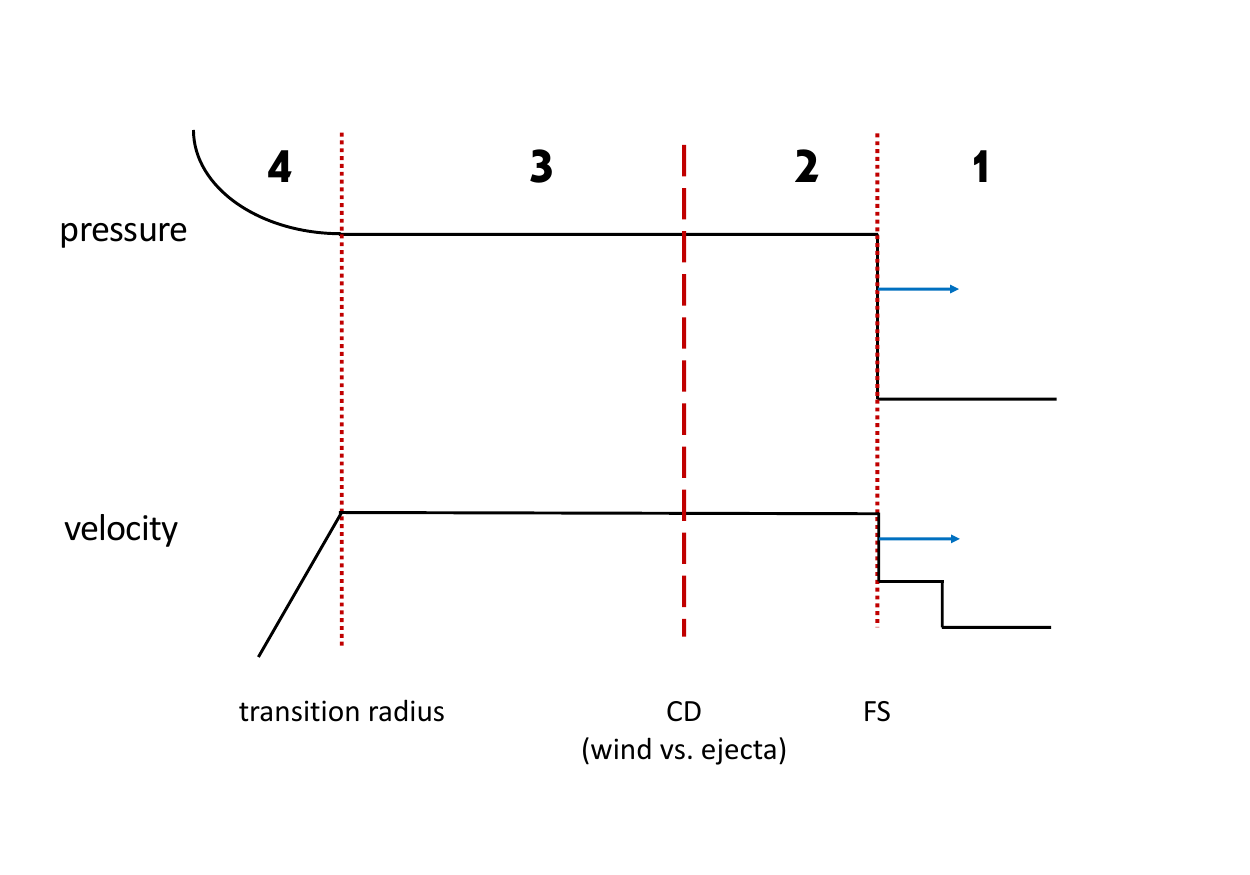}}\\
\resizebox{95mm}{!}{\includegraphics[]{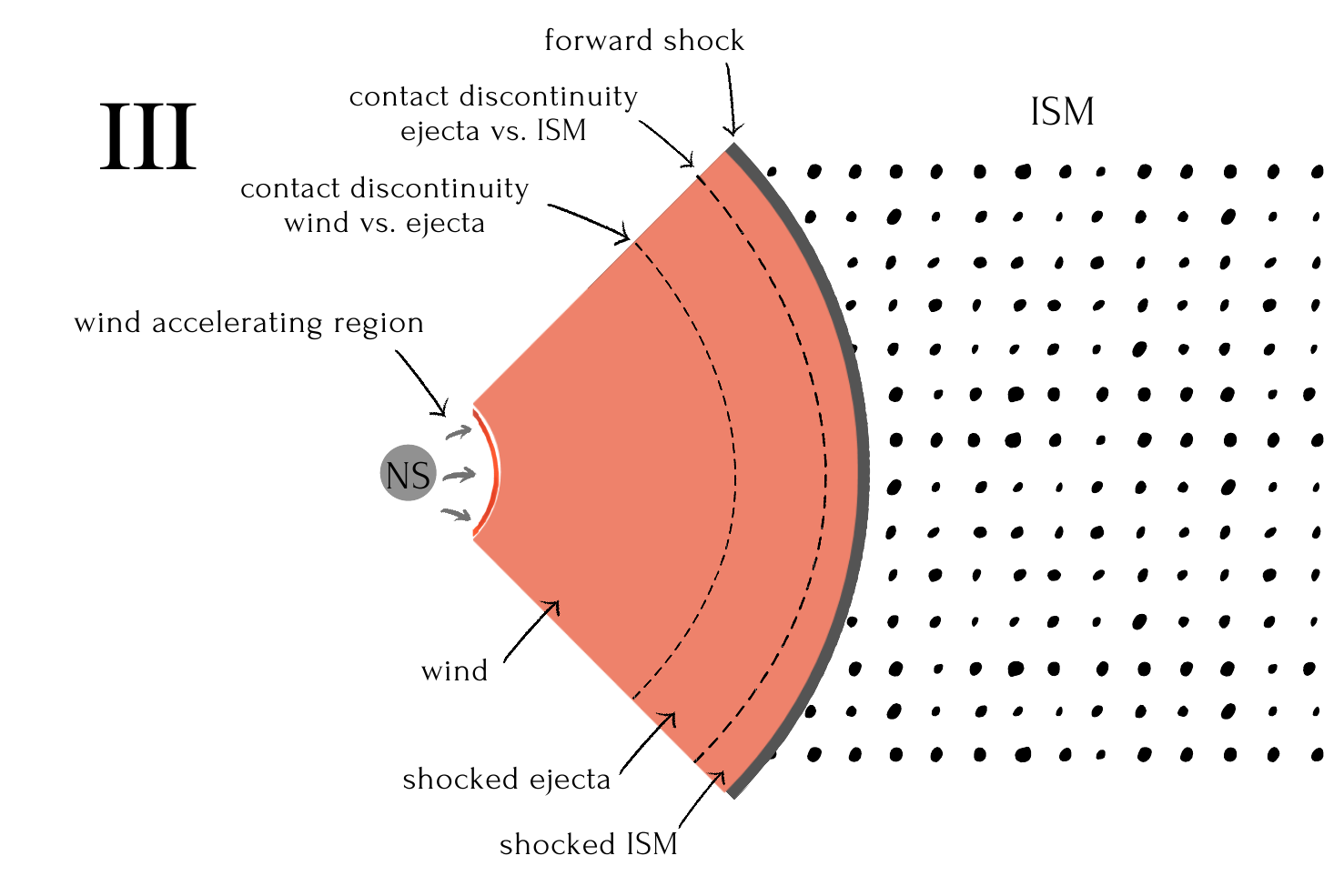}} &
\resizebox{90mm}{!}{\includegraphics[]{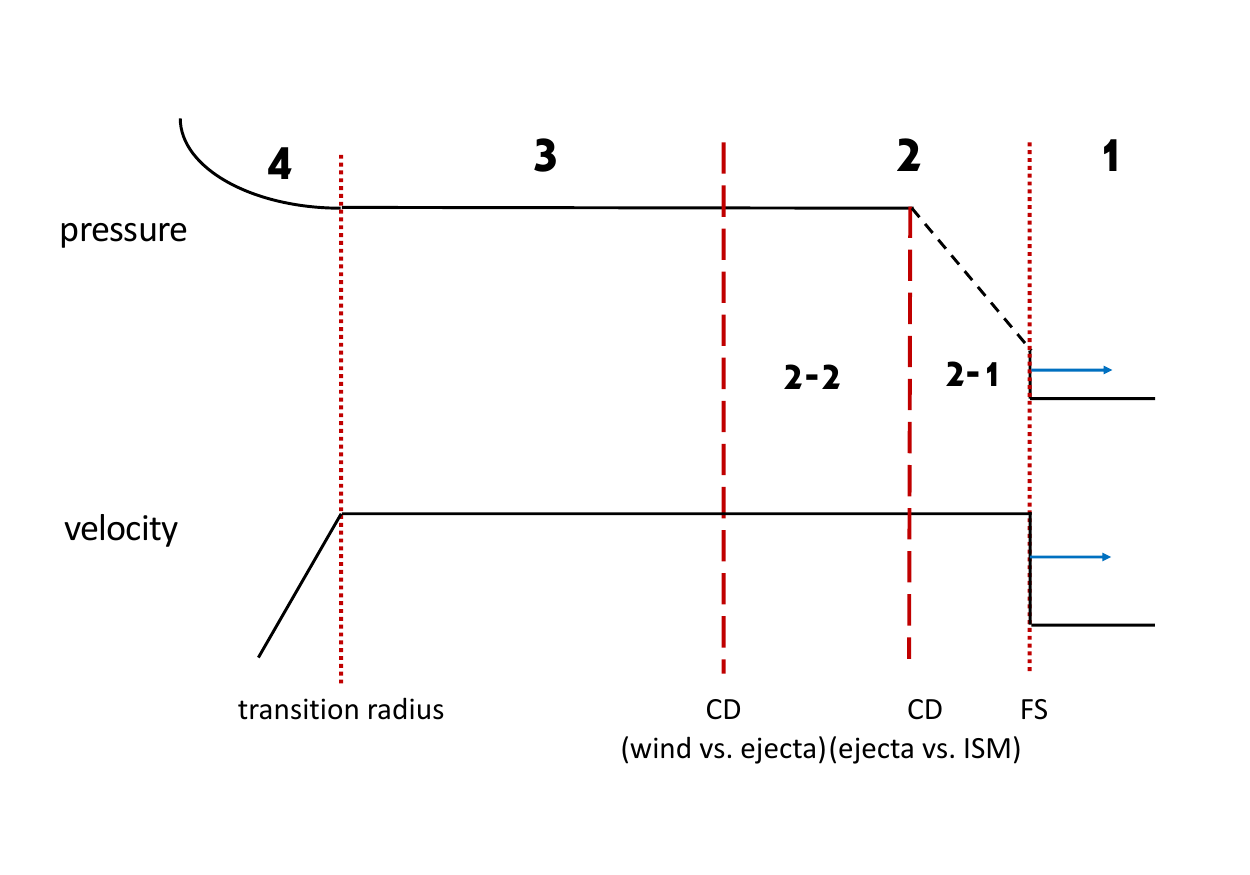}}
\end{tabular}
\caption{Schematic pictures for the energy injecting process from a central engine wind to the ejecta. The left panels are the cartoon illustrations for the shock systems, while the right panels show the profiles of pressure and velocity from the engine to the ISM, which are drawn under the assumption that the pressure is balanced in the blastwave. In our detailed treatment, the pressure balance assumption is released and a more accurate treatment is carried out using the mechanical model. The whole process can be divided into three stages. Stage \uppercase\expandafter{\romannumeral1}: A FS propagates into the ejecta while an RS propagates into the central engine wind. Stage \uppercase\expandafter{\romannumeral2}: The RS reaches the engine wind-acceleration region and forms a steady transition radius from the accelerating wind and the blastwave. Stage \uppercase\expandafter{\romannumeral3}: The FS breaks out from the ejecta and continues to propagating into the ISM. The whole system in all three stages is always divided into four regions, with the FS, RS (or the transition radius) and the contact discontinuity between the wind and ejecta as the boundaries. In stage III, although both the shocked ejecta and shocked ISM are matter-dominated and included as one part of the blastwave, to be clear, we further divide Region 2 into two sub-regions: 2-1 (shocked ISM) and 2-2 (shocked ejecta).}
\label{fig:schematic}
\end{figure*}
\begin{itemize}
    \item Stage \uppercase\expandafter{\romannumeral1}: An FS propagates outwards into the ejecta; An RS, in the meantime, propagates inwards to the central NS. In principle, there should be another pair of shocks at the interface between the ejecta and the interstellar medium (ISM). However, because the non-relativistic moving ejecta cannot excite a strong shock, this pair of shocks are ignored in the treatment. The entire system is then divided into four regions: (1) the unshocked ejecta; (2) the shocked ejecta; (3) the shocked wind; and (4) the unshocked wind. Assume that the unshocked wind is accelerated near the central engine. Beyond a certain radius, the velocity for the wind bulk motion is approximately treated as a constant. The pressure in the unshocked wind decreases with distance from the engine.
    \item Stage \uppercase\expandafter{\romannumeral2}: The FS keeps propagating in the ejecta; The RS vanishes when it reaches the accelerating zone. We set the radius where the Lorentz factor of the blastwave equals to that for the wind bulk motion as the inner boundary of the blastwave and the transition radius between the accelerating wind and the blastwave. The four regions of the whole system at this stage are defined as: (1) the unshocked ejecta; (2) the shocked ejecta; (3) the wind beyond the inner boundary; (4) the wind inside the inner boundary.
    \item Stage \uppercase\expandafter{\romannumeral3}: The FS breaks out from the ejecta and keep propagating into the ISM. The four regions for the whole system at this stage are defined as: (1) the unshocked ISM; (2) the shocked ejecta and ISM; (3) the wind beyond the inner boundary; (4) the wind inside the inner boundary. We also divide Region 2 into two sub-regions, 2-1 and 2-2, which represent the shocked ISM and shocked ejecta, respectively.
\end{itemize}

\section{Model}
\label{sec:model}
\subsection{Jump conditions for arbitrarily magnetized relativistic shocks}
\label{sec:jumpcondition}
Given the properties of the upstream for a shock (either the FS or RS), the properties of the downstream right behind the shock can be calculated through the jump conditions for an arbitrarily magnetized relativistic shock. We follow the derivations in \cite{zhangkobayashi05} and directly use the following useful equations
\begin{eqnarray}
&&u_{is}=u_{js}\Gamma_{ji}+(u_{js}^2+1)^{1/2}(\Gamma_{ji}^2-1)^{1/2},\\
&&\rho_j  = \frac{u_{is}}{u_{js}} \rho_i \\
&&p_j =\left(\Gamma_{ji}-1\right)\left(1-{\Gamma_{ji}+\frac{1}{2} u_{is}u_{js}}\sigma_i\right) \frac{n_j m_p c^2}{\hat{\gamma}-1} \\
&&B_j = \frac{u_{is}}{u_{js}} B_i. \label{eq:Bjump}
\end{eqnarray}
Here the subscript ``ji'' represents the physical quantity for region ``j'', in the rest frame of ``i''. The same applies for ``js'' and ``is'', with ``s'' stands for the ``shock''. 
The quantities with only one subscript are defined in their own rest frame, $u = \Gamma \beta$ represents the first component of the four-velocity vector, and $\hat{\gamma}$ represents the adiabatic index of the materials in the downstream, which can be estimated as
\begin{eqnarray}
\hat{\gamma} \approx \frac{4\Gamma_{ji} + 1}{3\Gamma_{ji}}.
\label{eq:gamma_hat}
\end{eqnarray}

To use the jump conditions above, the first step is to calculate $u_{js}$. Setting $x =u_{js}^2$, one needs to solve a third-order equation 
\begin{eqnarray}
Jx^3 + Kx^2 + Lx + M = 0,
\label{eq:u2s}
\end{eqnarray}
where 
\begin{eqnarray}
J &=& \hat{\gamma}(2-\hat{\gamma})(\Gamma_{ji}-1) + 2, \\
K &=& -(\Gamma_{ji}+1)[(2-\hat{\gamma})(\hat{\gamma}\Gamma_{ji}^2+1)+\hat{\gamma}(\hat{\
\Gamma}-1)\Gamma_{ji}]\sigma_i \nonumber \\
&&-(\Gamma_{ji}-1)[\hat{\gamma}(2-\Gamma)(\Gamma_{ji}^2 - 2)+(2\Gamma_{ji}+3)] \\
L&=& (\Gamma_{ji} + 1)[\hat{\gamma}(1-\frac{\hat{\gamma}}{ 4})(\Gamma_{ji}^2-1)+1]\sigma_i^2 \nonumber \\
&&+ (\Gamma_{ji}^2-1)[2\Gamma_{ji}-(2-\hat{\gamma})(\hat{\gamma}\Gamma_{ji}-1)]\sigma_i \nonumber \\
&&+(\Gamma_{ji}-1)(\Gamma_{ji}-1)^2(\hat{\gamma}-1)^2 \\
M &=& -(\Gamma_{ji}-1)(\Gamma_{ji}+1)^2(2-\hat{\gamma})^2 \frac{\sigma_i^2}{4}.
\end{eqnarray}

In a shock pair system, once the Lorentz factor of the blastwave is known, $\Gamma_{ji}$ for either the FS or RS can be calculated so that $u_{js}$ can be obtained numerically. Substituting the $u_{js}$ value together with the quantities in the upstreams into the jump conditions, the quantities in the downstreams can be calculated. The magnetic pressure is defined as $p_b = B^2/(8 \pi)$. Thus, the jump condition for magnetic pressure is written as $p_{b,j}=(u_{is}^2/u_{js}^2)p_{b,i}$.
The dimensionless speed of shock propagation in the lab frame can be calculated from $u_{js}$, which reads
\begin{eqnarray}
\beta_s = \frac{\beta_j + \beta_{js}}{1 + \beta_j \beta_{js}},
\label{eq:beta_s}
\end{eqnarray}
where $\beta_{js} = u_{js}/\sqrt{(1 + u_{js}^2)}$.

\subsection{Mechanical model for bastwaves}
\label{sec:mechanical}
%We define Region 2 and Region 3 together as the ``blatwave''. 

According to the definition in Section \ref{sec:description}, the blastwave includes Region 2 and Region 3.
In the simple analytical treatments of GRB problems, the total pressure (including gas pressure and magnetic pressure) is assumed to be balanced in the blastwave \citep{saripiran95,zhangkobayashi05}. An obvious flaw for this assumption is that energy is not conserved \citep{beloborodov06,uhm11}. Since we are exploring the energy injecting efficiency, which is directly related to the energy conservation issue, the pressure gradient in the blastwave must be introduced. Indeed, the existence of a pressure gradient in the GRB problem has been verified by 1D MHD simulations \citep{mimica09}. On the other hand, based on the hydrodynamical \citep{kobayashi00} and MHD simulation (see Section \ref{sec:simulation}) results, the gradient of bulk motion velocity in the blastwave is very small. Therefore, one may still assume a constant bulk motion Lorentz factor in space within the blastwave. Based on this assumption, \cite{beloborodov06} proposed a mechanical model for non-magnetized fluids to fix the energy non-conservation issue in the GRB problems. Instead of considering each fluid element, they treated the blastwave as a whole and study the evolution of the integrated quantities over radius. \cite{uhm11} applied this model to the case of a long-lasting wind interacting with the ISM and found that the energy conservation law is still well satisfied. Recently, motivated by treating more realistic central engine energy injection problems invoking a Poynting-flux-dominated central engine wind,  \cite{ai21} generalized the mechanical model to arbitrarily magnetized blastwaves based on the jump conditions for magnetized relativistic shocks \citep{zhangkobayashi05}. In this paper, we adopt a slightly modified version of the magnetized mechanical model to study the engine-fed kilonova (mergernova) problem.

For an arbitrary fluid element in the magnetized relativistic blastwave under spherical geometry, one can write the equation of continuity 
\begin{eqnarray}
\frac{\partial}{\partial t}(\Gamma \rho) + \frac{\partial}{\partial r} (\Gamma \rho \beta c) + \frac{2\Gamma \rho \beta c}{r} = 0,
\label{eq:mass1}
\end{eqnarray}
the momentum equation  
\begin{eqnarray}
&&\frac{1}{c}\frac{\partial}{\partial t}(\Gamma^2 h \beta) + \frac{1}{4\pi c}\frac{\partial} {\partial t}(\Gamma^2 \beta B^2) +\frac{1}{r^2}\frac{\partial}{\partial r}(r^2 \Gamma^2 h \beta^2) + \frac{\partial p}{\partial r} \nonumber\\
&& + \frac{1}{8\pi}\frac{\partial}{\partial r}[(1+\beta^2)\Gamma^2 B^2] + \frac{1}{4\pi r}(1+\beta^2)\Gamma^2 B^2 = 0,
\label{eq:momentum1}
\end{eqnarray}
and the energy equation
\begin{eqnarray}
&&\frac{\partial}{\partial t}(\Gamma^2 h) - \frac{\partial}{\partial t}p + \frac{1}{8\pi}\frac{\partial}{\partial t}[(1+\beta^2)\Gamma^2 B^2] + \frac{1}{r^2} \frac{\partial}{\partial r}(r^2 \Gamma^2 h \beta c) \nonumber \\
&&+ \frac{c}{4\pi r^2}\frac{\partial}{\partial r}( r^2\beta \Gamma^2 B^2) = 0,
\label{eq:energy1}
\end{eqnarray}
where $h = \rho c^2 + \hat{\gamma}p/(\hat{\gamma}-1)$ is the specific enthalpy. In these equations, $\rho$, $p$, $h$ and $B$ are defined in the rest frame of the fluid, while other quantities are defined in the lab frame. $\beta$ is the bulk motion velocity of the blastwave.

Since the profiles of these quantities in the blastwave are unknown, we only focus on its evolution as a whole, thus take the integrals of Equations \ref{eq:mass1}, \ref{eq:momentum1} and \ref{eq:energy1}, from the RS ($r_r$) to the FS ($r_f$). In our previous magnetized mechanical model \citep{ai21}, integrated quantities over radius were defined (i.e. $\Sigma = \int_{r_r}^{r_f} \rho dr$, etc.) similar to the non-magnetized mechanical model \citep{beloborodov06,uhm11}. Such a treatment is only accurate when the blastwave is thin, with all the quantities approximately centered at the contact discontinuity. If the central engine wind is highly magnetized and the forward shock travels slowly in a dense medium (as required by the engine-fed kilonova problem we are treating in this paper), Region 3 is relatively very thick. Without the profile of $B$, one cannot determine the effective radius of the integrated $B$ (marked as ${\cal B}$ in \cite{ai21}). To be more accurate, here we modify the mechanical model by defining the integrated quantities over the {\em volume} instead of over the radius in the blastwave, which read 
\begin{eqnarray}
\Sigma_{\rm sph} &=& \int_{r_r}^{r_f} 4\pi r^2 \rho dr, \\
P_{\rm sph} &=& \int_{r_r}^{r_f}  4\pi r^2 p dr, \\
H_{\rm sph} &=& \int_{r_r}^{r_f}  4\pi r^2 h dr, \\
{\cal B}_{\rm sph} &=& \int_{r_r}^{r_f}  4\pi r^2 B^2 dr.
\end{eqnarray}
When multiplying the equations for an arbitrary fluid element with $4\pi r^2 dr$ and take their integrals, respectively, Equations \ref{eq:mass1}, \ref{eq:momentum1} and \ref{eq:energy1} now become
\begin{flalign}
&\beta \frac{d}{dr_d}(\Gamma \Sigma_{\rm sph}) - 4\pi \Gamma [\rho_r r_r^2 (\beta-\beta_r) - \rho_f r_f^2 (\beta_f - \beta)] = 0,&
\label{eq:mass_int}
\end{flalign}

\begin{flalign}
&\beta \frac{d}{dr_d}(\Gamma^2 \beta H_{\rm sph}) + \frac{\beta}{4\pi} \frac{d}{dr_d}(\Gamma^2 \beta {\cal B}_{\rm sph})& \nonumber\\
&- 4\pi \Gamma^2 \beta [h_r r_r^2 (\beta -\beta_r) + h_f r_f^2(\beta_f - \beta)]& \nonumber \\
&+ \Gamma^2 \beta (B_r^2 r_r^2 \beta_r - B_f^2 r_f^2 \beta_f) + \frac{1}{2} (1+\beta^2)\Gamma^2 (B_f^2 r_f^2 - B_r^2 r_r^2)&  \nonumber \\
&+ 4\pi (p_f r_f^2 - p_r r_r^2) - 8\pi \int_{r_r}^{r_f} rp dr = 0,&
\label{eq:momentum_int}
\end{flalign}

\begin{flalign}
&\beta \frac{d}{dr_d} (\Gamma^2 H_{\rm sph}) - \beta \frac{d}{dr_d} P_{\rm sph} + \frac{\beta}{8\pi} \frac{d}{dr_d} [\Gamma^2 (1+\beta^2){\cal B}_{\rm sph}]& \nonumber \\
&- 4\pi \Gamma^2 [h_r r_r^2(\beta-\beta_r) + h_f r_f^2 (\beta_f - \beta)]& \nonumber \\
&- 4\pi [p_r r_r^2 \beta_r - p_f r_f^2 \beta_f] + \frac{1}{2} (1+\beta^2)\Gamma^2 (B_r^2 r_r^2 \beta_r - B_f^2 r_f^2 \beta_f)& \nonumber \\
&+ \Gamma^2 \beta [B_f^2 r_f^2 - B_r^2 r_r^2] = 0.&
\label{eq:energy_int}
\end{flalign}
The subscripts ``f'' and ``r'' represent the quantities right behind the FS the RS, respectively, in the rest frame of the blastwave ($\beta_f$ and $\beta_r$ are, however, defined in the lab frame), and $r_d$ is the radius of contact discontinuity. In our calculation, the identity
\begin{flalign}
\int_{r_r(t)}^{r_f(t)} \frac{\partial}{\partial t}f(t,r) dr = \frac{d}{dt} \left[\int_{r_r}^{r_f} f(t,r) dr \right]
+ c[f_r \beta_r - f_f \beta_f]
\end{flalign}
is used, where $f(r,t)$ can be any function of $r$ and $t$. Besides, $dr_d = \beta c dt$ is used to relate the time with the distance of the ejecta (represented by the contact discontinuity) from the central engine.
 
For the highly magnetized Region 4, one has $p_r \ll p_f$. The integrated thermal pressure is mainly accumulated from the FS and stored in Region 2. Hence, the effective adiabatic index in the entire blastwave may be approximately adopted for the shocked ejecta. In the ejecta, radiation pressure is dominant, so that $\hat{\gamma}_{\rm eff} = 4/3$ can be adopted (see Appendix \ref{sec:dominant_pressure}). The equation of state for the gas in the blastwave can be written as
\begin{eqnarray}
\frac{dH_{\rm sph}}{dr_d} = \frac{d\Sigma_{\rm sph}}{dr_d} c^2 + \frac{\hat{\gamma}_{\rm eff}}{{\hat{\gamma}}_{\rm eff}-1}\frac{dP_{\rm sph}}{dr_d}.
\label{eq:EOS_int}
\end{eqnarray}

We can also assume that the magnetic fields are confined below $r_d$. If magnetic dissipation is ignored, the magnetic energy cannot be converted to gas pressure or radiation pressure directly. Define $P_{\rm sph}^{\prime} = \int_{r_r}^{r_f} p dV^{\prime} = \Gamma P_{\rm sph}$ as the integrated gas pressure in the rest frame of the blastwave. Its evolution can be then expressed as\footnote{To be concise, here we ignore the release of internal energy (pressure) through thermal radiation, because radiative cooling is not significant compared to shock heating. See section \ref{sec:thermal_radiation} for details.} 
\begin{eqnarray}
\frac{dP_{\rm sph}^{\prime}}{dr_d} = \frac{dP_{\rm sph}^{\prime}}{dr_d}\bigg|_{\rm rs} + \frac{dP_{\rm sph}^{\prime}}{dr_d}\bigg|_{\rm fs} + \frac{dP_{\rm sph}^{\prime}}{dr_d}\bigg|_{\rm exp}, 
\label{eq:dP_sph}
\end{eqnarray}
where the three terms on the right hand side of the equation, from left to right, represent the accumulation
from the RS and FS, and energy release due via the $p dV$ work, respectively. Specifically, these terms read

\begin{eqnarray}
\frac{dP_{\rm sph}^{\prime}}{dr_d}\bigg|_{\rm rs}&=&4\pi r_r^2 \frac{\beta - \beta_r}{\beta} p_r \label{eq:dP_rs}\\
\frac{dP_{\rm sph}^{\prime}}{dr_d}\bigg|_{\rm fs}&=&4\pi r_f^2 \frac{\beta_f - \beta}{\beta} p_f \label{eq:dP_fs}\\
\frac{dP_{\rm sph}^{\prime}}{dr_d}\bigg|_{\rm exp} &=& -(\hat{\gamma}_{\rm eff}-1)\frac{P_{\rm sph}}{V}\frac{dV^{\prime}}{dr_d}  \nonumber \\
&=& -(\hat{\gamma}_{\rm eff}-1)\left(P_{\rm sph} \frac{d\Gamma}{dr_d} + \frac{2\Gamma P_{\rm sph}}{r_p}\right) \label{eq:dP_exp}
\end{eqnarray}

In our problem, assuming that the gas pressure is effectively centered at the contact discontinuity, one can adopt $r_p = r_d$. The last term in Equation \ref{eq:momentum_int} can be also  simplified as 
\begin{eqnarray}
-8\pi \int_{r_r}^{r_f} r p dr = -\frac{2P_{\rm sph}}{r_d}.
\label{eq:r_p}
\end{eqnarray}
Now, combining Equations \ref{eq:mass_int}, \ref{eq:momentum_int}, \ref{eq:energy_int}, \ref{eq:EOS_int}, and \ref{eq:dP_sph}, one can solve the evolution of each quantity. 

The RS vanishes when it reaches the transition radius around $r_{\rm acc}$. Take this radius as the inner boundary of the blastwave. Energy keeps being injected into the blastwave from this boundary. Instead of solving the jump conditions, every quantity is continuous across the boundary. Therefore, in Stages II and III, the subscript ``r'' loses its original meaning of the ``RS''. In order to maintain a unified notion, hereafter we still use the subscript ``r'' to generally denote the inner boundary, which is the transition radius between the wind and the blastwave. With the acceleration of the blastwave, the position of the inner boundary may increase slightly, but would stay within the accelerating region. After the RS vanishes, we always set $\Gamma_w = \Gamma_{\rm w,acc} = \Gamma$ and $r_r = r_{\rm acc}$.

\section{MHD simulation with \emph{Athena++}}
\label{sec:simulation}
In this section, we present a 1D relativistic MHD simulation for Riemann problems with \emph{Athena++} \citep{stone20}, to verify the basic assumptions in Section \ref{sec:model} (e.g. constant bulk motion velocity and pressure gradient) and the validity of the mechanical model for magnetized relativistic blastwaves. However, the real problem is challenging to simulate due to the following two reasons: (1) The NS wind is highly magnetized, and (2) The density ratio between the massive ejecta and other parts is extremely high. Therefore, instead of using the real values and real ratios, we only use a set of less extreme parameters to perform the simulation.

\begin{figure*}
\begin{tabular}{lll}
\resizebox{55mm}{!}{\includegraphics[]{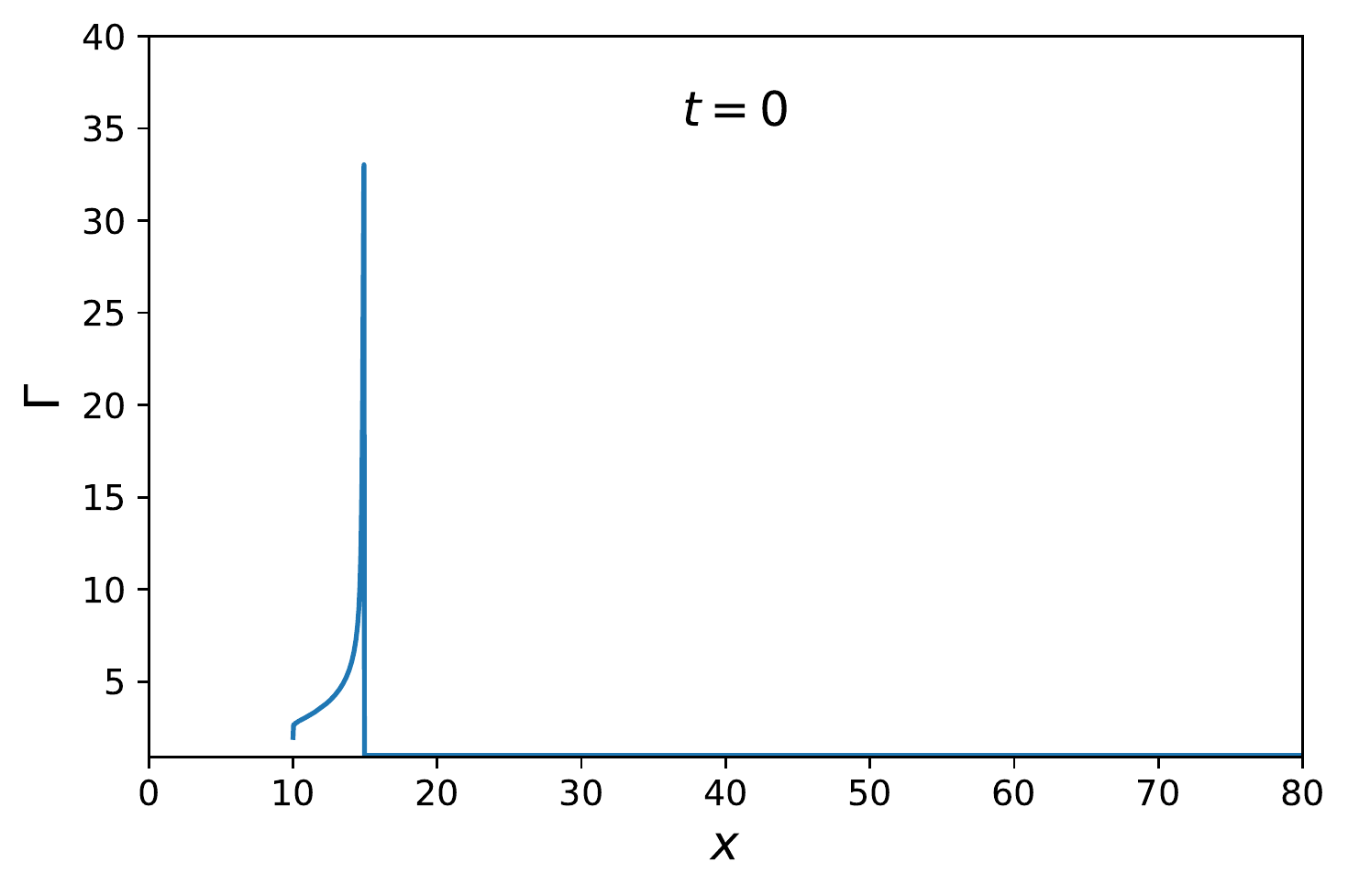}} &
\resizebox{55mm}{!}{\includegraphics[]{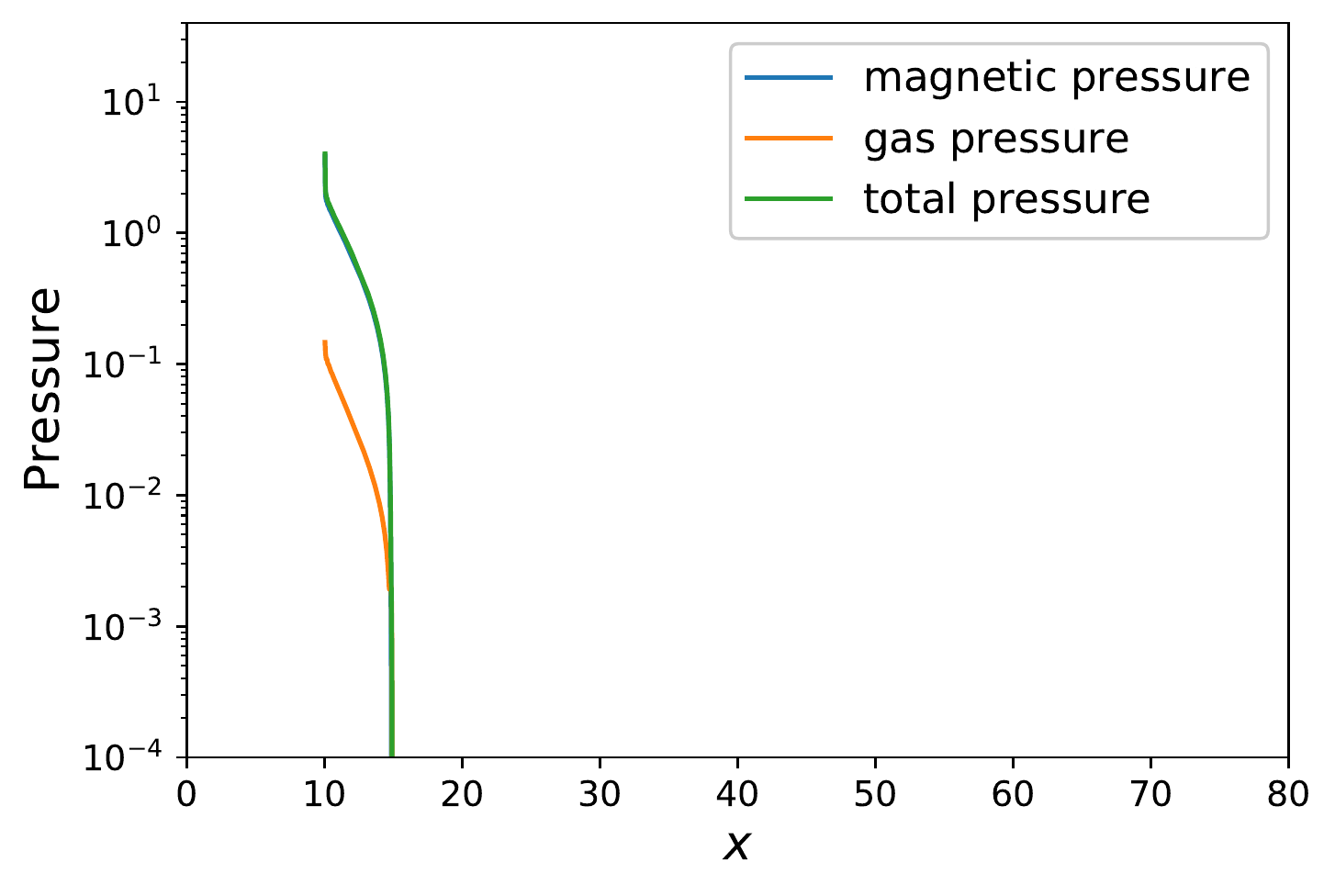}} &
\resizebox{55mm}{!}{\includegraphics[]{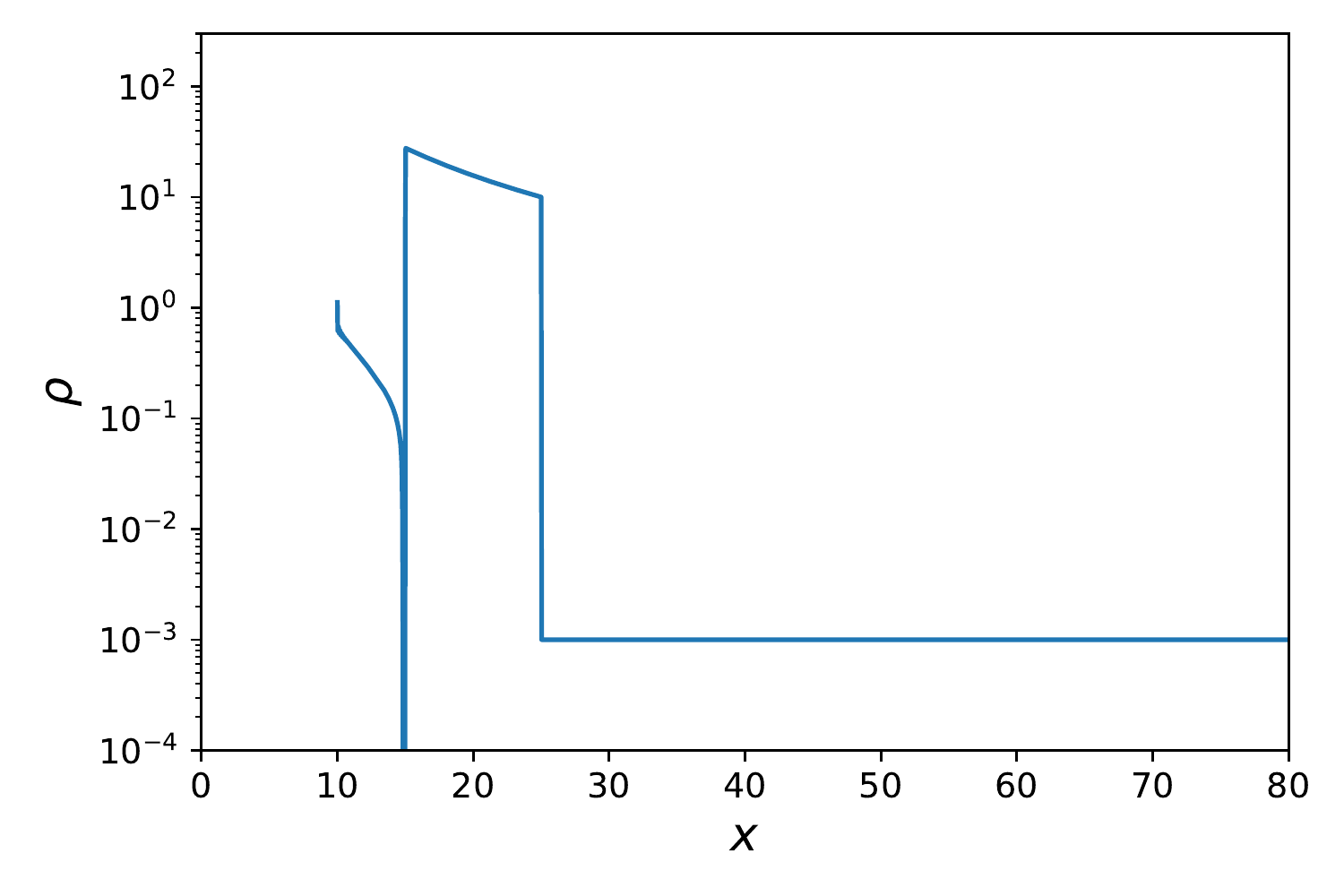}} \\
\resizebox{55mm}{!}{\includegraphics[]{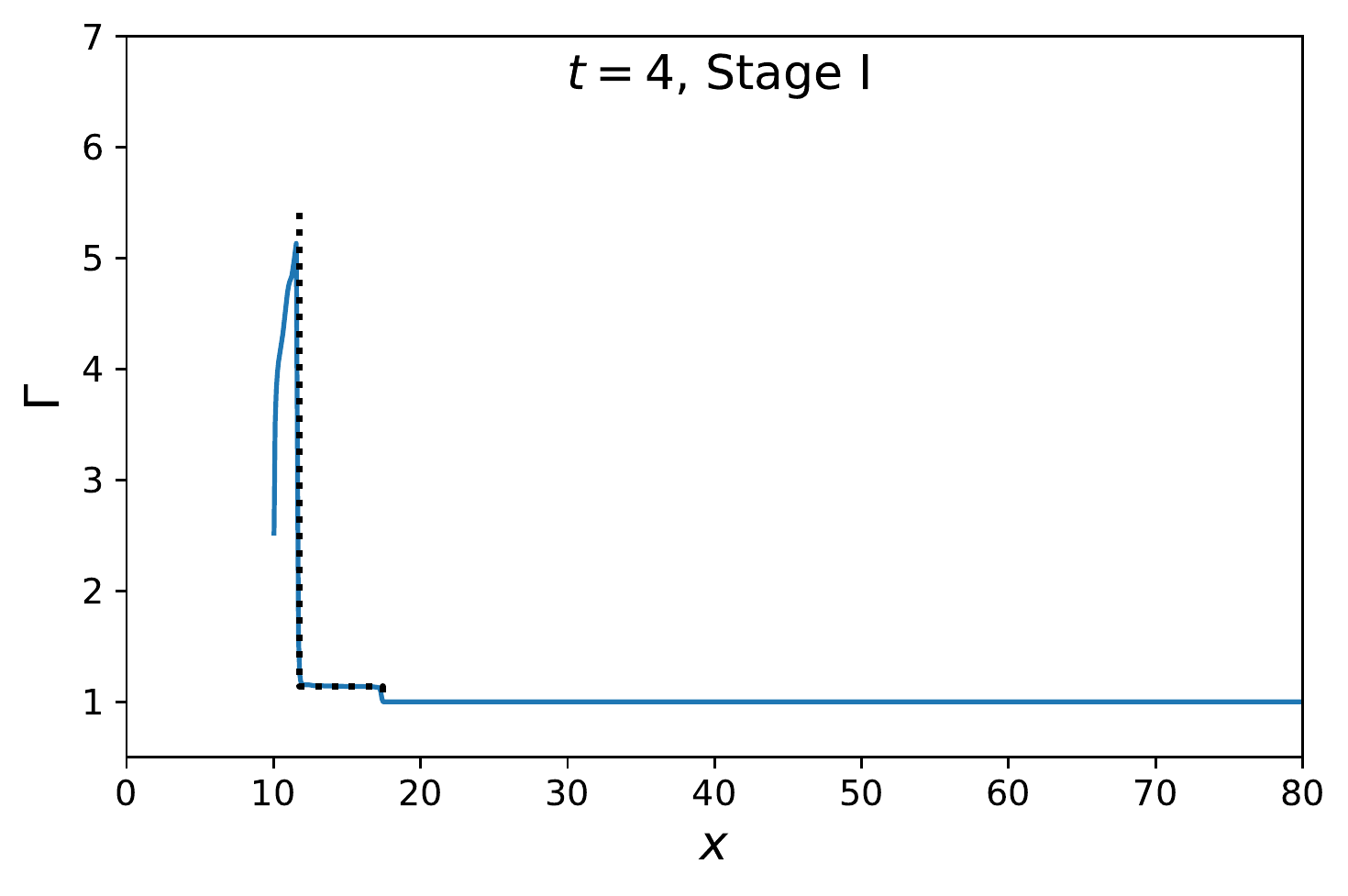}} &
\resizebox{55mm}{!}{\includegraphics[]{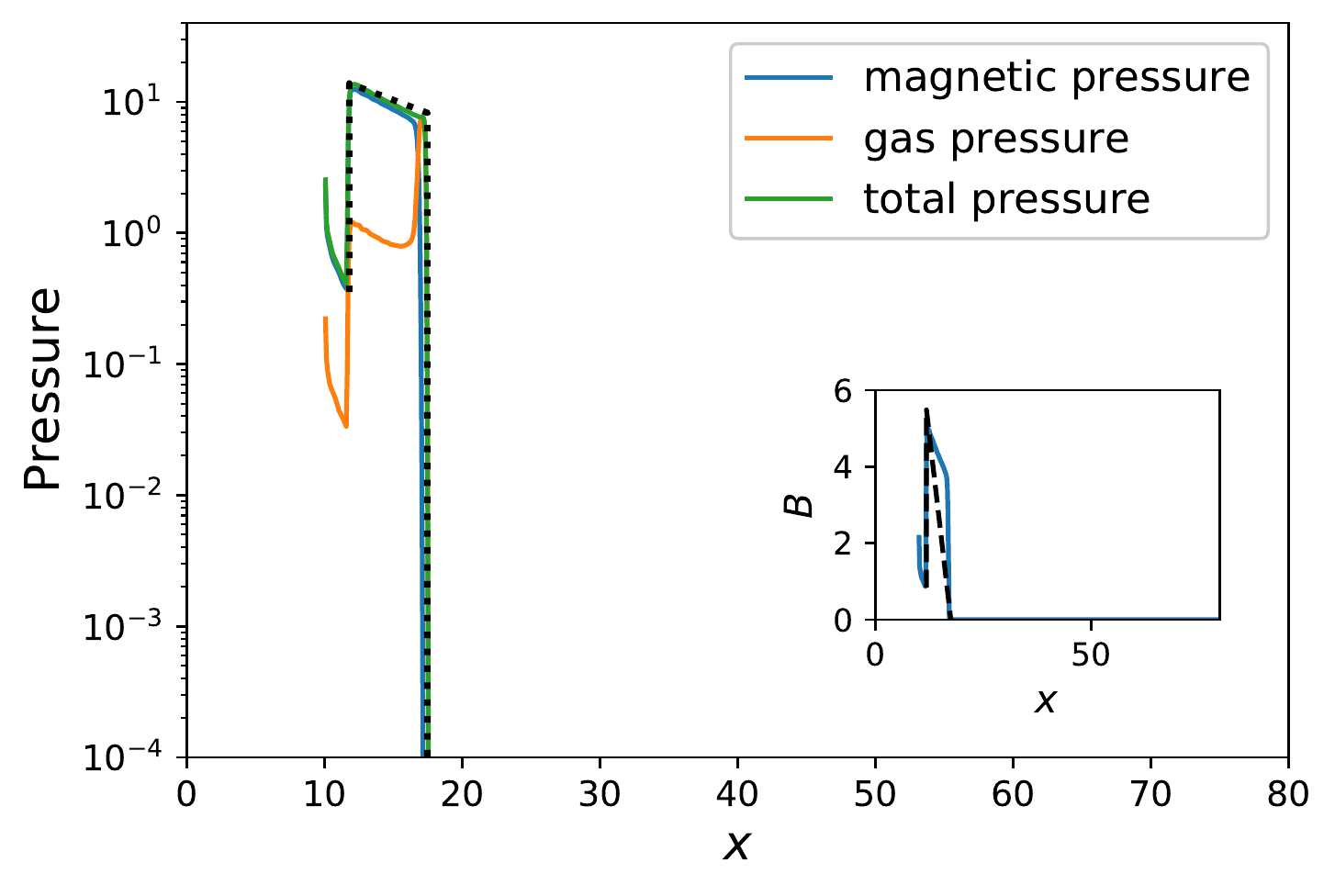}} &
\resizebox{55mm}{!}{\includegraphics[]{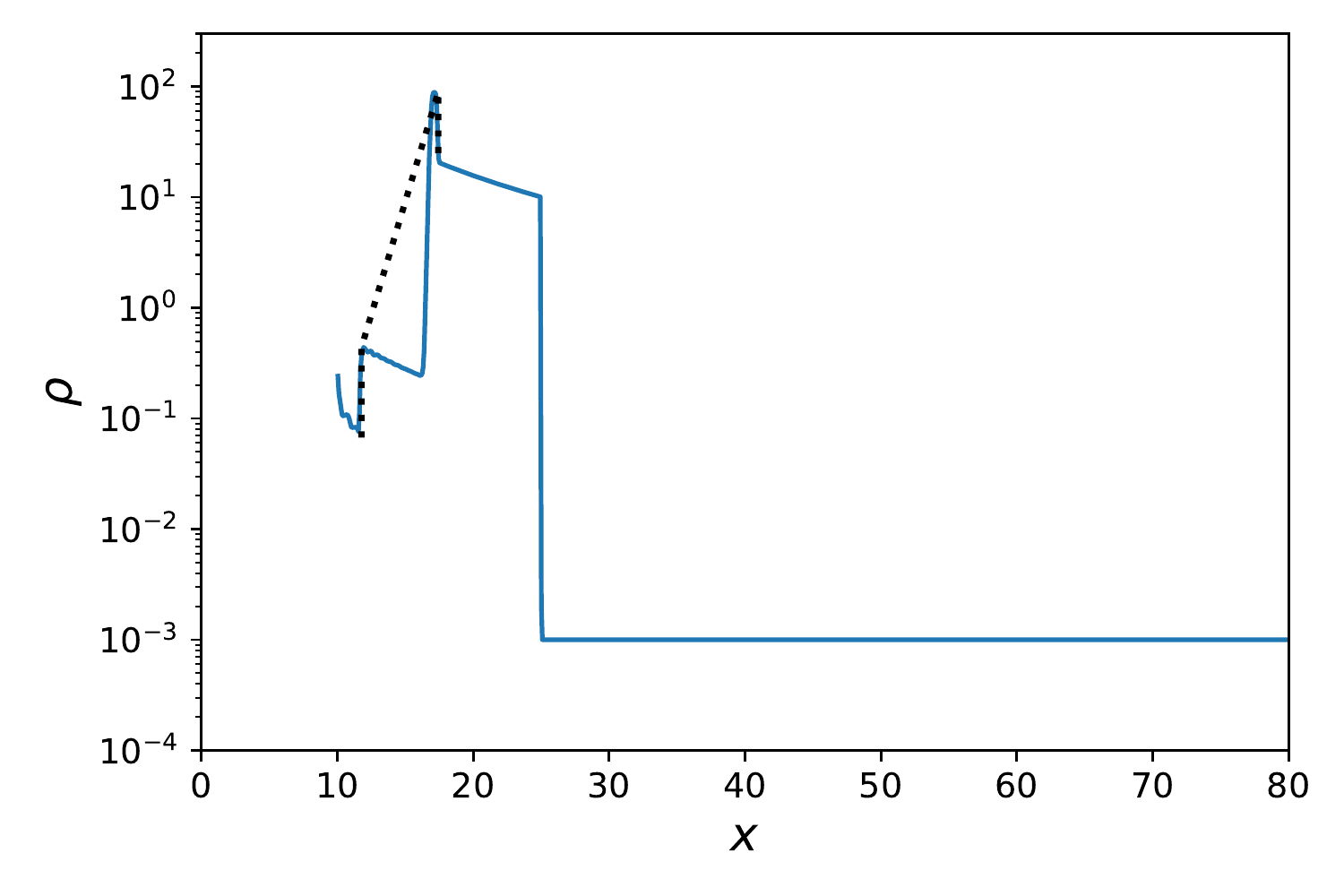}} \\
\resizebox{55mm}{!}{\includegraphics[]{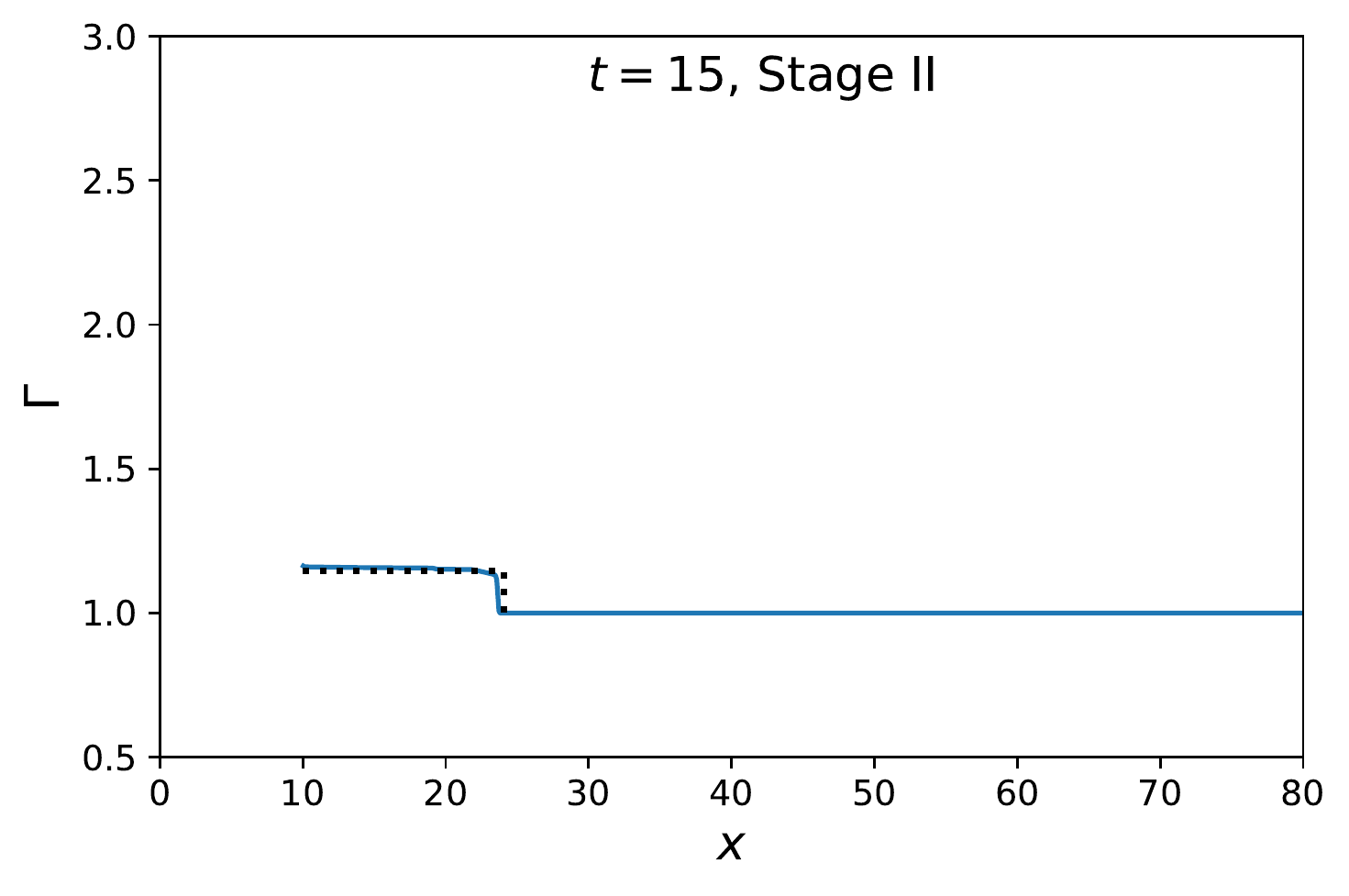}} &
\resizebox{55mm}{!}{\includegraphics[]{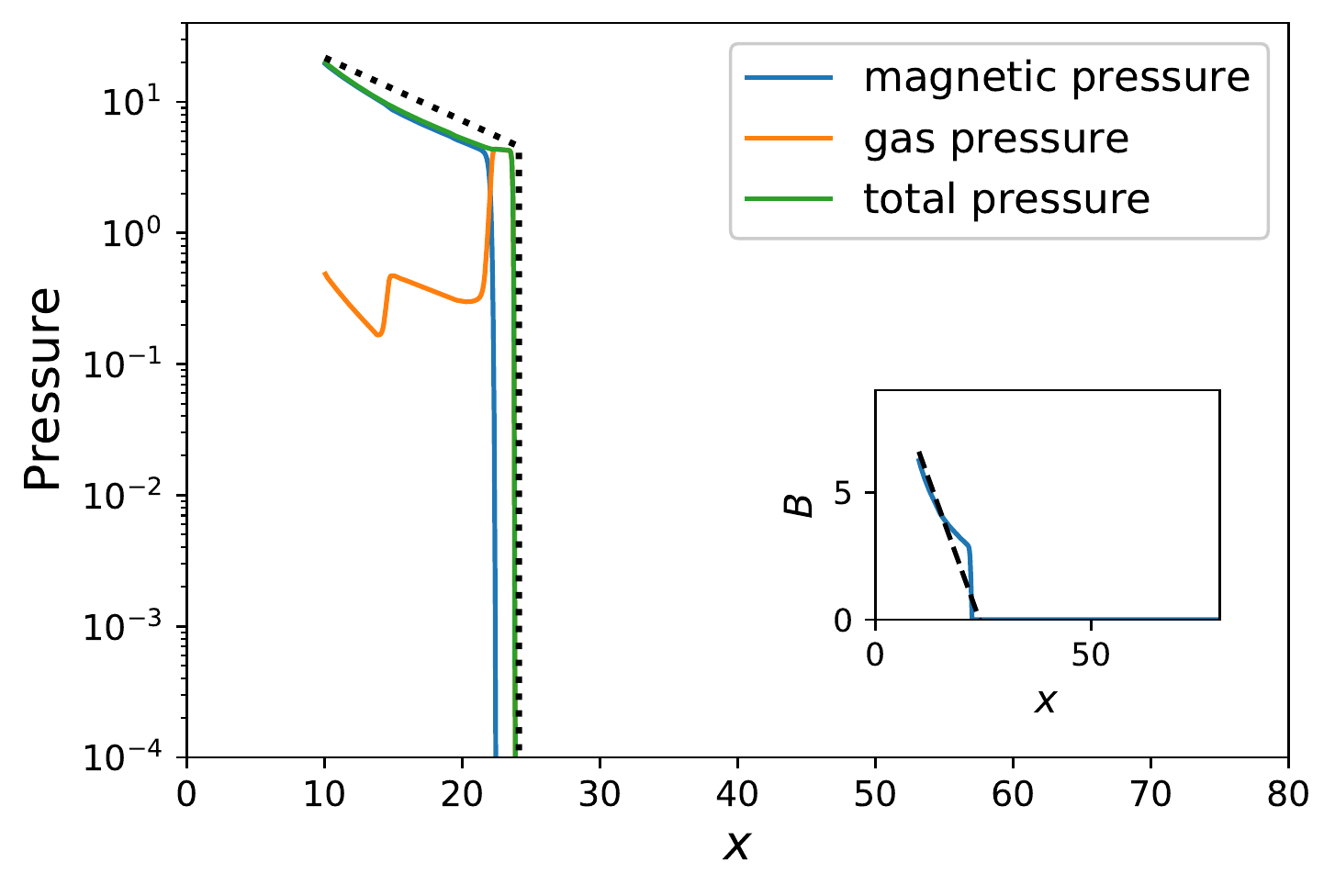}} &
\resizebox{55mm}{!}{\includegraphics[]{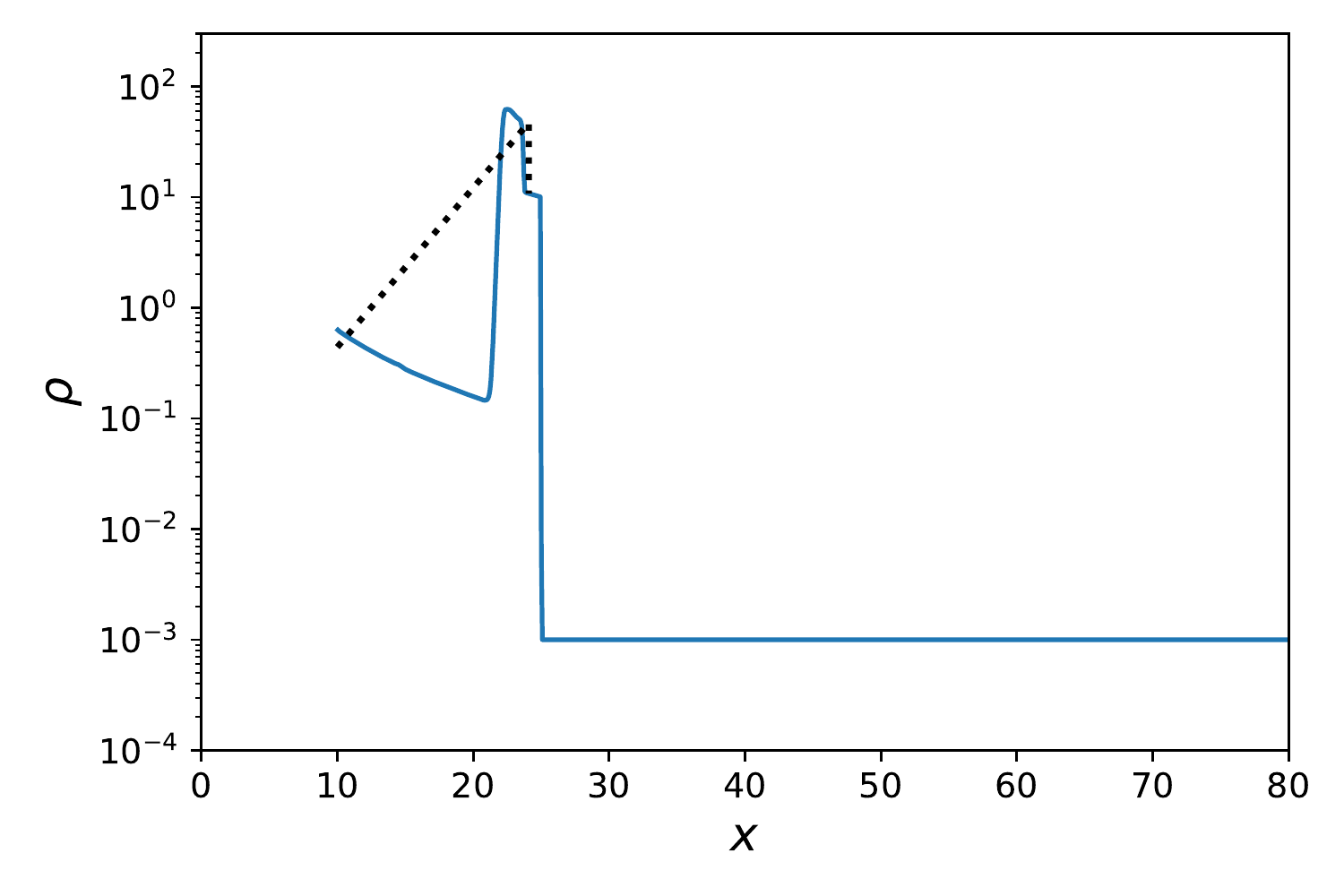}} \\
\resizebox{55mm}{!}{\includegraphics[]{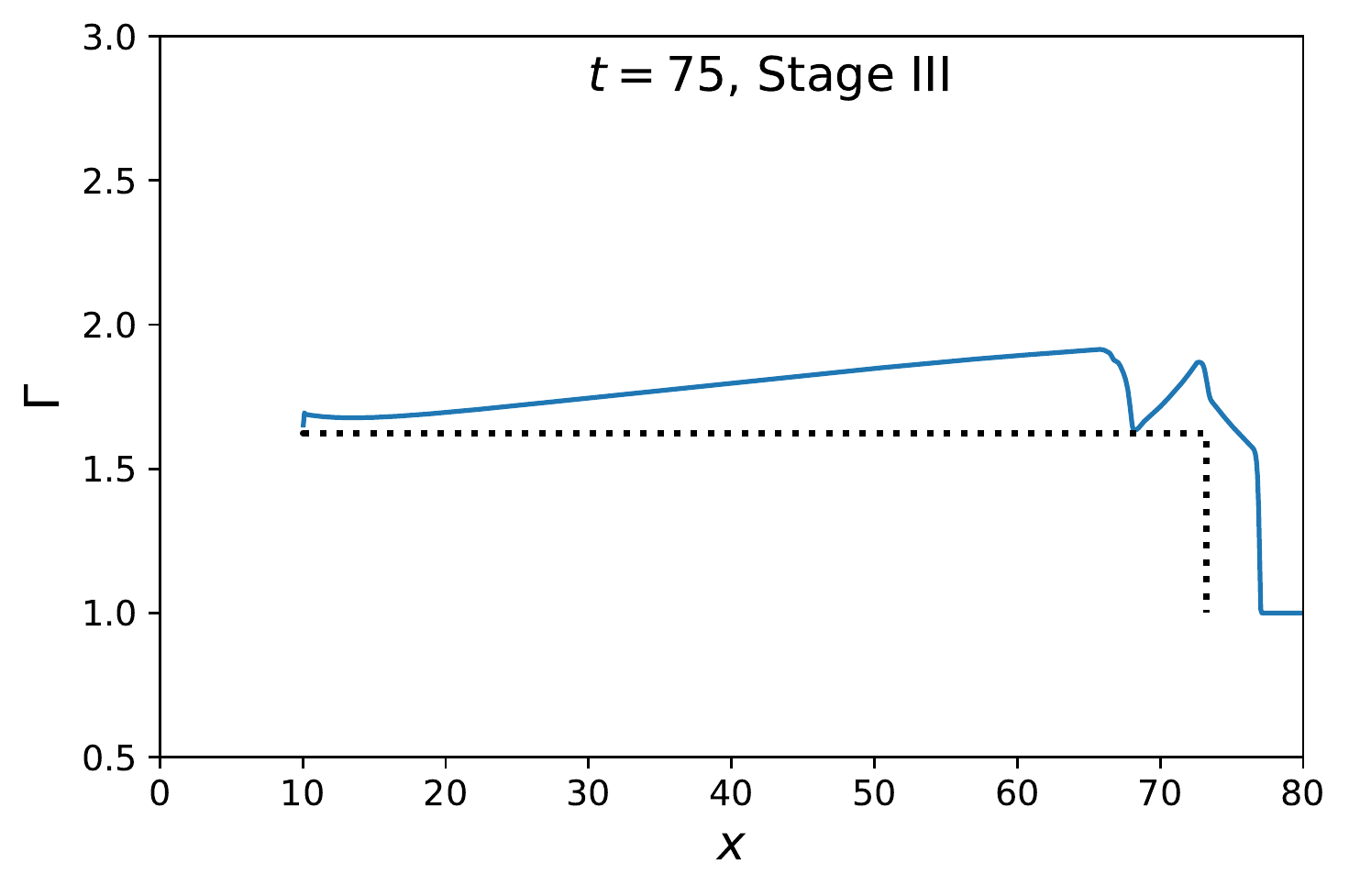}} &
\resizebox{55mm}{!}{\includegraphics[]{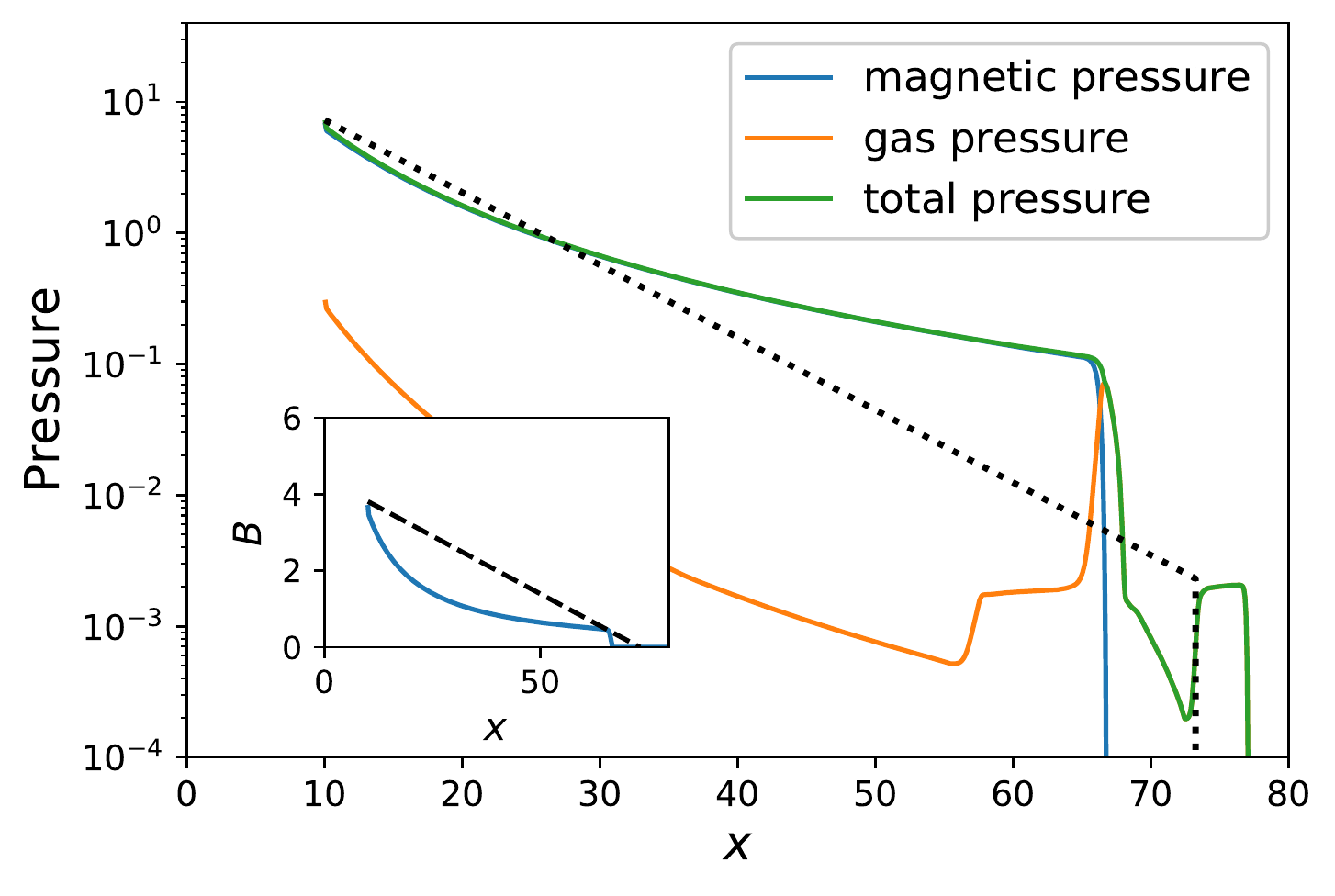}} &
\resizebox{55mm}{!}{\includegraphics[]{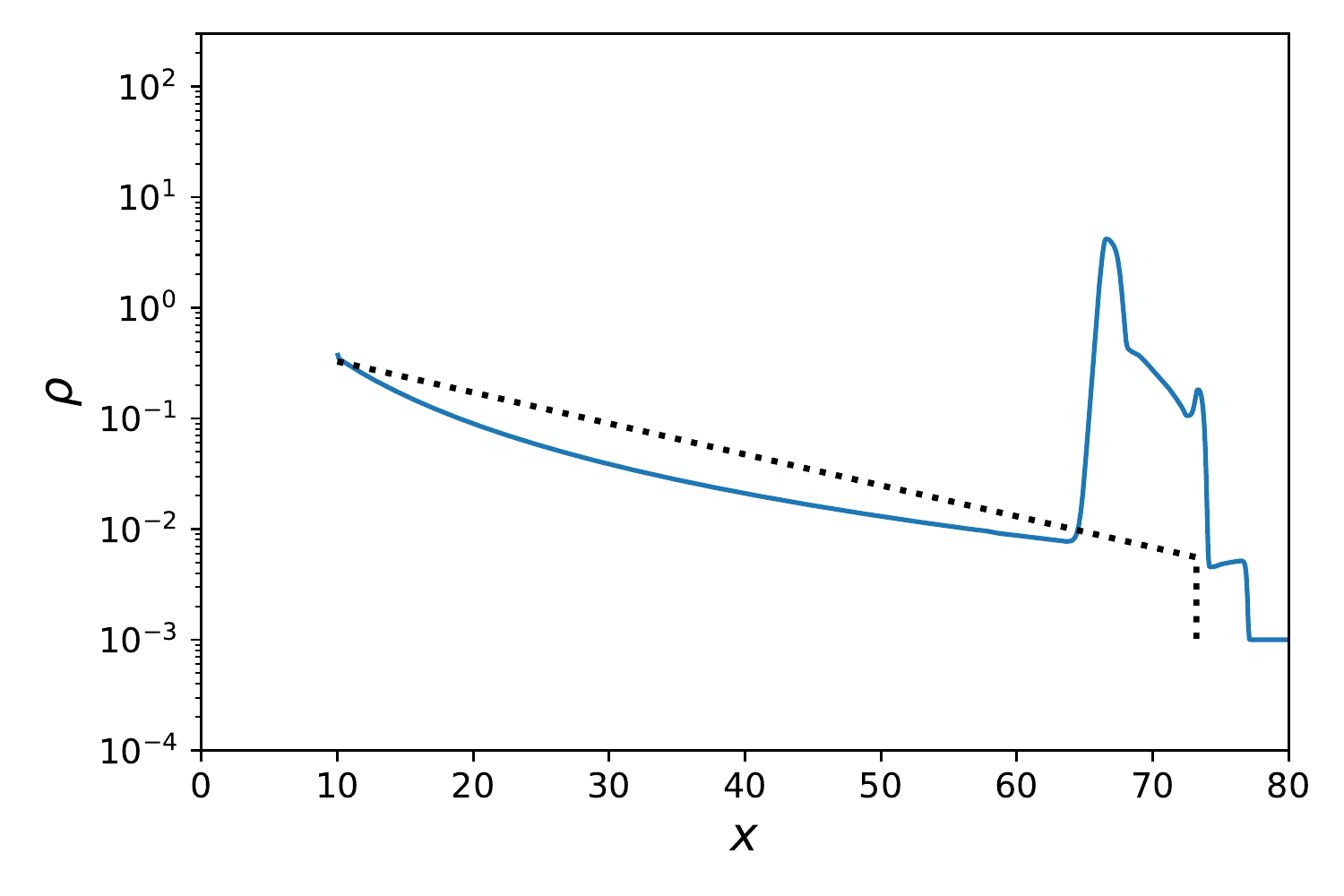}} \\
\resizebox{55mm}{!}{\includegraphics[]{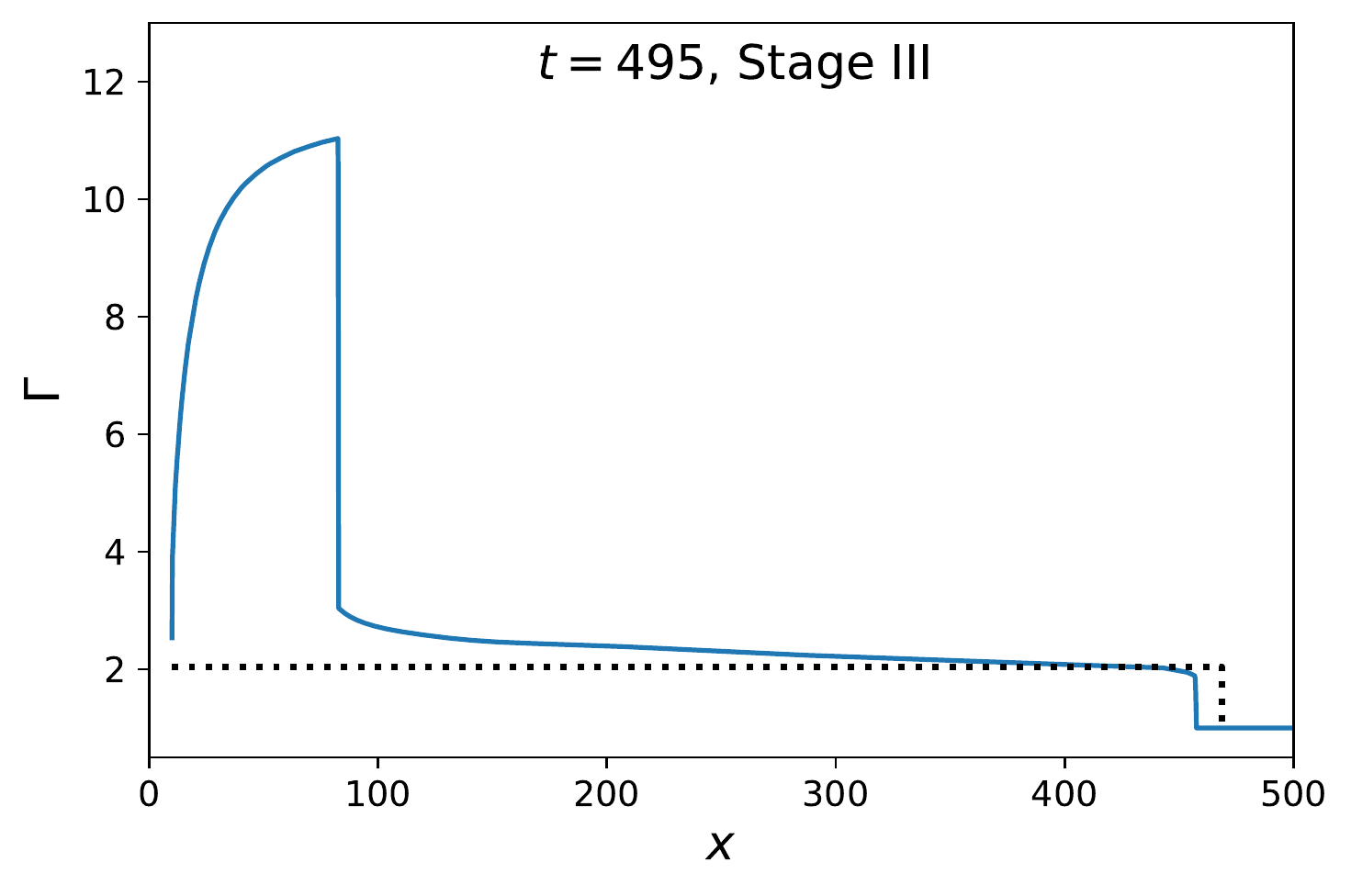}} &
\resizebox{55mm}{!}{\includegraphics[]{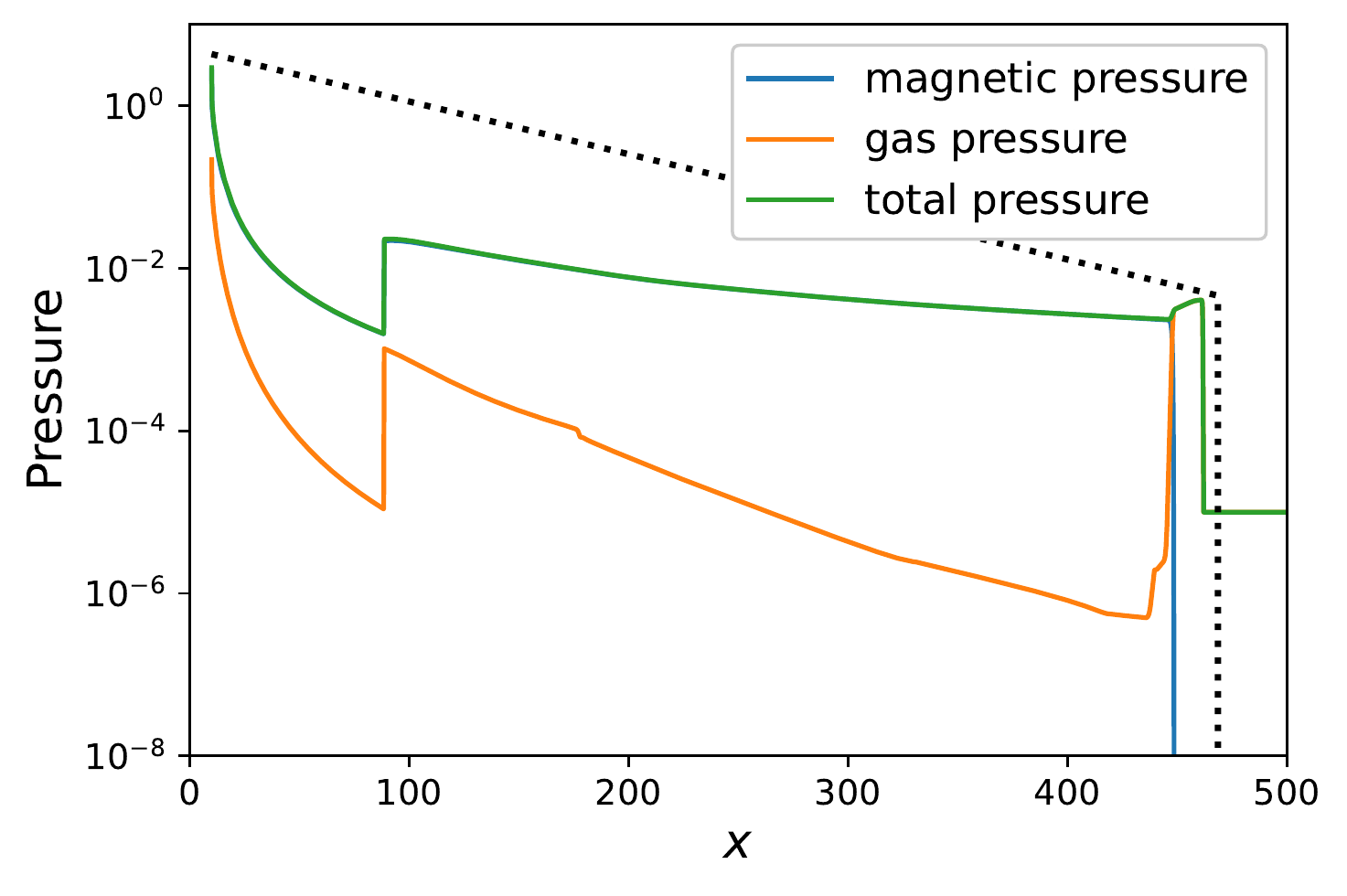}} &
\resizebox{55mm}{!}{\includegraphics[]{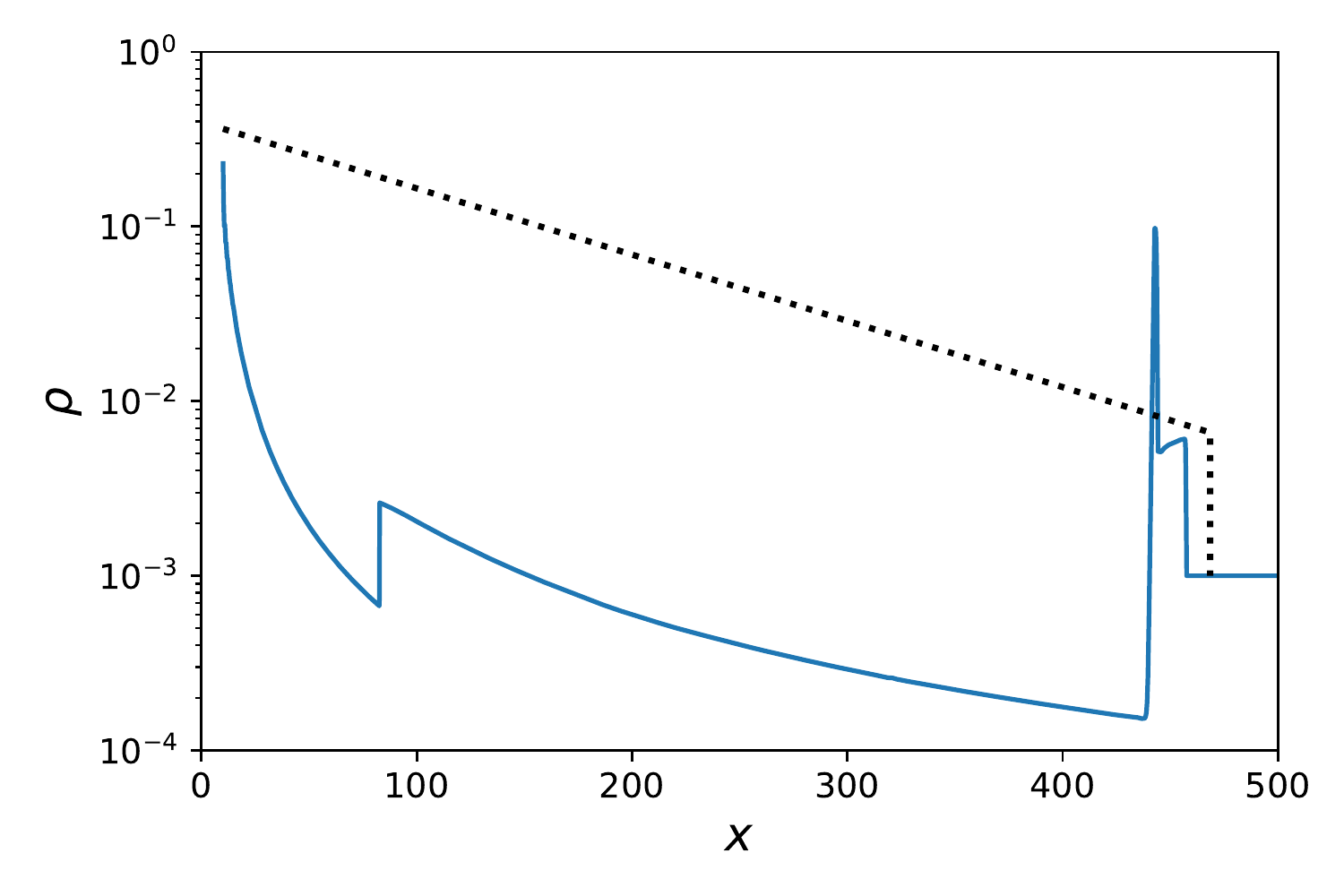}}
\end{tabular}
\caption{The profiles of the bulk motion Lorentz factor, pressures and density at different stages. The profiles in different stages (at different times) are shown in different rows while the profiles of different quantities are shown in different columns. The dotted black lines stand for the results from the mechanical model while the solid lines stand for the results from the MHD simulation. Since the mechanical model deals with integrated quatities only, the profiles from the mechanical model cannot be derived and we simply connect the values right behind the FS and the RS (inner boundary).}
\label{fig:athena}
\end{figure*}

\begin{figure}
\resizebox{80mm}{!}{\includegraphics[]{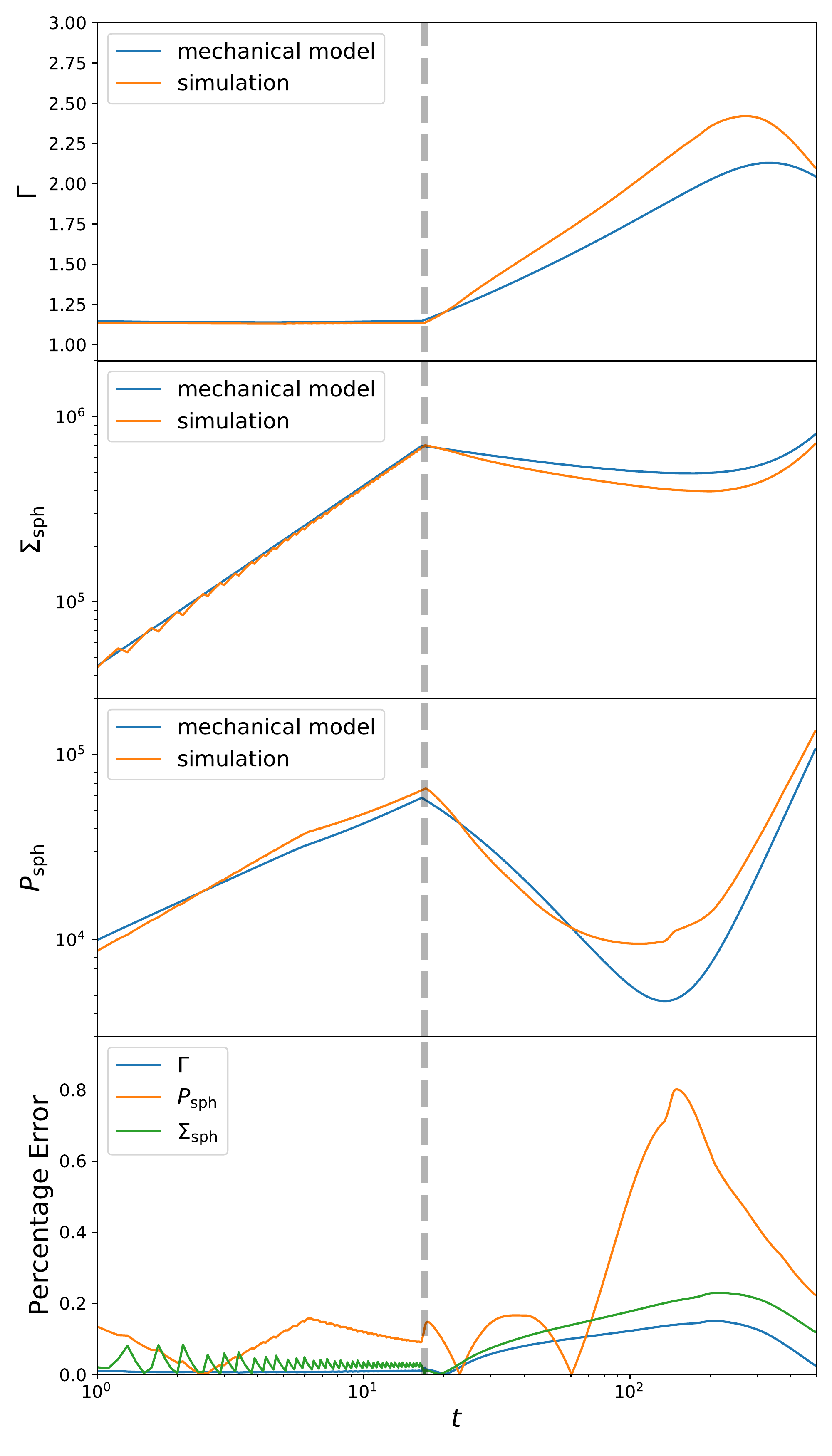}}
\caption{The comparison between the evolution of some (integrated) physical quantities from the MHD simulation and the mechanical model. The bottom panel shows the percentage errors of the mechanical model, assuming that the results from the simulation are true values. The vertical dashed lines represent the time when the FS crosses the ejecta.}
\label{fig:compare}
\end{figure}

The simulation is set up in the spherical geometry, with special relativity enabled. In code units, the light speed is normalized to $c=1$. The magnetization parameter is defined as $\sigma = B^2/\rho$, where $B$ is the magnetic strength in the rest frame. %which is assumed to be along the $\phi$ direction.
In our setup, the ejecta is initially at rest, located between $x = 15$ and $x = 25$, with $\rho_{\rm rj} \propto x^{-2}$ and $\rho_{\rm ej} (x = 25) = 10$; The uniform ISM fills the region where $x > 25$, with $\rho_{\rm ISM} = 0.001$. We set the inner boundary at x=10 from which the wind is blown out. Highly magnetized materials are set up in the ghost cells within $x = 10$ with $\rho_{\rm gho} = 1.0$, $b_{\rm gho} = 10.0$, and $v = 0.1$. $b_{\rm gho}$ is along the $\phi$ direction. However, the huge pressure gradient between the ghost cells and the first active cell from the inner boundary ($x = 10$) accelerates the wind. Therefore, the wind particles are blown out with a bulk motion velocity greater than $v = 0.1$. Then, the wind material would continue to be accelerated until it contacts with the ejecta. Initially, the ejecta and the ISM is set to be cold ($p=0$). Also, the gas pressure can be ignored in the highly magnetized wind. We set the first wind-ejecta contact time as $t = 0$, after which we evolve the whole system by both conducting the MHD simulation with Athena++ and numerically solving the differential equations in the mechanical model, respectively. The adiabatic index $\hat{\gamma}$ depends on the particles' random motion velocities. Suppose the random motion velocities of particles in the upstream are non-relativistic, $\hat{\gamma} = 5/3$ should be adopted there. In the downstream, $\hat{\gamma}$ can be estimated with the strength of the shocks from Equation \ref{eq:gamma_hat}. The strength of the FS is low ($\Gamma_{21} \sim 1$). Hence, in Region 2, $\hat{\gamma} \approx 5/3$ can also be adopted. Region 3 is dominated by magnetic fields, where the influence of the $\hat{\gamma}$ value is not significant. Therefore, we can effectively set $\hat{\gamma} \approx 5/3$ in the whole system. 

We simulate the evolution of the system over the three stages defined in section \ref{sec:description}, and present the results in Figure \ref{fig:athena}. The results from our mechanical model are shown with dashed lines for a comparison. In Stage I, we observed an FS-RS pair. The blastwave between the FS and RS has a nearly constant bulk motion velocity in space, which verifies the basic assumption of the mechanical model. Also, the pressures in the blastwave have obvious gradients, suggesting the pressure balance assumption is not suitable for our problem. In Stage II, the RS reaches the inner boundary and then vanishes, which is also consistent with our prediction. In Stage III, the FS crosses the ejecta and continues to propagate in the ISM. Meanwhile, since the shocked ejecta is hot, a rarefaction wave forms and starts to propagate inwards through the ejecta. In this stage, the bulk motion velocity of the materials in the blastwave is no longer a constant in space, violating the basic assumption for the mechanical model. However, after the rarefaction wave crosses the whole blastwave, the velocity gradient can be eliminated. At late times, when the pressure of the shocked wind decreases to a certain level, a new FS would form in the previously shocked wind. The pressure of the shocked wind near the inner boundary decreases because the blastwave has been significantly accelerated (to a relativistic speed). Even if the new FS can catch up with and heat up the ejecta again, the heating effect should not be comparable with that of the first FS. This is because the relative Lorentz factor of the wind and the ejecta must be smaller than that of the first interaction. Therefore, we treat the newly formed FS just as a substructure. As can be seen from Figure \ref{fig:athena}, in Stage I, Stage II and late-time Stage III, the results from our mechanical model presents a good approximation to the MHD simulation results. In all stages, the profiles of gas pressure and density indicate that matter and internal energy are mainly stored in Region 2, thus Equation \ref{eq:r_p} is also valid.

For the same parameter set, we also solve the mechanical model together with the jump conditions to find the positions of the shocks, as well as the physical quantities near the shocks, at different times. In Stage I, since the analytical expression of the unshocked wind profile is unknown, we just apply the simulated one directly to our semi-analytical model. The luminosity of the wind can be calculated with any cell in Region 4. In Stage II and III, the physical quantities at the inner boundary can be calculated from the the wind luminosity together with the bulk motion velocity of the blastwave, which is equal to the bulk motion velocity of the wind (see Section \ref{sec:region4} for details). The properties of Region 1 can be easily obtained based on the position of the FS. Since the mechanical model deals with the integrated quantities only and cannot provide the profiles of physical quantities in the blastwave, the black dashed lines in Figure \ref{fig:athena} are just simple connections between the values right behind the FS and the RS (inner boundary). 
%To compare the results from the mechanical model with that from the MHD simulations, we manually connect the values right behind the shocks and compare with the results from the mechanical model with dotted lines in Figure \ref{fig:athena}. 

As can be seen, both the FS/RS positions and the physical quantity values near the FS/RS derived from the mechanical model are consistent with the simulation results in all the stages except the early phase of Stage III. Please note that the profiles shown with the dotted lines in Figure 2 have no physical meaning, because we cannot really derive the profile of each physical quantity from the mechanical model. Therefore, it does not matter although the density does not match well in Stage III. In the early phase of Stage III, the position of the FS does not match well either, because the velocity gradient introduced by the inward propagating rarefaction wave in the front of the ejecta can excite a stronger FS in the ISM than that derived from the mechanical model. Next, we compare the (integrated) quantities directly related to the ejecta between the results from the simulation and the mechanical model. Using the simulated values as the true values, the percentage errors for the mechanical model can be calculated. As shown in Figure \ref{fig:compare}, the percentage errors remain lower than $\sim 15\%$ before the FS crossing, after which deviation appears mainly due to the rarefaction wave discussed above. At very late times, the deviation become smaller again as expected. 

\section{Dynamics with realistic parameters}
\label{sec:dynamics}
In the following, we first discuss the physics in different regions in detail in a realistic astrophysical environment for the engine-fed kilonova problem. Then we calculate the dynamical evolution of the blastwave with our mechanical model.
\subsection{Region 1}
\label{sec:region1}
Before the FS breaks out (Stage I and II), the unshocked ejecta serves as Region 1. Suppose that the unshocked ejecta is wind-like (i.e. $\rho \propto r^{-2}$). The density of Region 1 in its rest frame can be calculated as
\begin{eqnarray}
\rho_1 = \frac{M_{\rm ej}}{4\pi r_f^2 \Delta_{\rm ej,0} \Gamma_{\rm ej}},
\end{eqnarray}
where $M_{\rm ej}$ and $\Delta_{\rm ej,0}$ represent the rest mass and initial thickness of the ejecta in the lab frame, respectively, $\Gamma_1 = \Gamma_{\rm ej}$ is the Lorentz factor for the ejecta bulk motion, and $r_f$ is the position of the FS in the lab frame. In our treatment, the dimensionless velocity of the ejecta is adopted as a fiducial value of $\beta_{\rm ej}= v_{\rm ej}/c = 0.2$,
%where $c$ stands for the speed of light,
so $\Gamma_{\rm ej} \sim 1.02$ which is close to unity. Since the ejecta is assumed to be cold and non-magnetized, we set the thermal pressure $p_1 = 0$ and the magnetic field strength $B_1 = 0$ in the problem set-up.

The FS crosses the ejecta when $\int_0^{t_\times} (\beta_f - \beta_{\rm ej}) c dt = \Delta_{\rm ej}$, where $\beta_f$ and $t$ stand for the FS dimensionless velocity and the time in the lab frame. After $t_\times$, Region 1 becomes the unshocked ISM, in which we have $\Gamma_1=1$, $p_1 = 0$ and $B_1 = 0$. The density of Region 1 is $\rho_1 \sim n_{\rm ISM} m_p$, where $n_{\rm ISM}$ is the number density of the ISM, which is adopted to have a fiducial value of $n_{\rm ISM} = 0.1 {\rm cm^{-3}}$. 

In our calculations, the initial position of the ejecta is assumed to be at $r = 10^{10}{\rm cm}$, with a width in the lab frame $\Delta_{\rm ej,0} = 10^{12}{\rm cm}$. 

\subsection{Region 4}
\label{sec:region4}
Let us consider a magnetar central engine. It should spin down via  magnetic dipole radiation and, probably in the early phase, also by gravitational radiation. In the following, we assume that the magnetic dipole radiation dominates the spindown. Hence, the luminosity of the magnetar wind can be written as the spindown luminosity $L_{\rm sd}$, which would stay roughly constant before the spindown timescale $t_{\rm sd} = E_{\rm rot}/L_{\rm sd,0}$, where $E_{\rm rot} = (1/2)I\Omega_0^2 \approx 2\times 10^{52}{\rm erg}~(I/10^{45} {\rm g ~ cm^2}) (P_{0}/1 {\rm ms})^{-2}$ is the initial rotation energy of the magnetar, $I$, $\Omega_0$ and $P_0$ represent the moment of inertia, initial spin angular velocity and initial spin period of the magnetar. In principle, significant energy injection before spindown can happen both in stage II and stage III.

For the unshocked wind, in its rest frame, the density can be calculated as
\begin{eqnarray}
\rho_4 = \frac{L_{\rm sd}}{4\pi r^2 \Gamma_w^2 \beta_w c^3 (1+\sigma_{wL})}.
\label{eq:rhowind}
\end{eqnarray}
where $\sigma_{wL} = (2\Gamma^2-1)/(2\Gamma^2) \sigma_w$ stands for the ratio between the magnetic energy and the mass energy in the lab frame.
According to the definition of the magnetization parameter, the magnetic field can be calculated as
\begin{eqnarray}
B_4 &=& (4\pi \rho_4 c^2 \sigma_w)^{1/2} \nonumber \\
&=& \Gamma_w^{-1} r^{-1} \left(\frac{L_{\rm sd}}{ \beta_w c}\right)^{1/2}. \left(\frac{\sigma_w}{1+\sigma_{wL}}\right)^{1/2}
\label{eq:Bwind}
\end{eqnarray}
The criterion for RS formation (Equation \ref{eq:RScriteria}) is equivalent to
\begin{flalign}
\frac{\sigma_w}{1+\sigma_{wL}} < &9.0\times 10^7& \nonumber \\
&\beta_w \left(\frac{M_{\rm ej}}{10^{-3}M_{\odot}}\right)\left(\frac{\Gamma_w}{10^3}\right)^2 \left(\frac{L_{\rm sd}}{10^{49}{\rm erg/s}}\right)^{-1} \left(\frac{\Delta_{\rm ej,0}}{10^{12}{\rm cm}}\right)^{-1}&,
\end{flalign}
which is always satisfied. Hence, an RS is formed upon the interaction between the central engine wind and the ejecta. 

We consider that the wind is generated and accelerated near the light cylinder of the millisecond magnetar ($r_{\rm acc} \sim 10^7 {\rm cm}$), beyond which we set both $\Gamma_w$ and $\sigma_w$ as constants. Treating $r_{\rm acc}$ as the transition radius in stage II and stage III, the relation $\Gamma_{\rm w,acc} = \Gamma$ should be satisfied, where $\Gamma_{\rm w,acc}$ and $\Gamma$ represent the Lorentz factor for the bulk motion of the unshocked wind at $r_{\rm acc}$ and the bulk motion of the blastwave, respectively. Since the wind is highly magnetized, the value of $\rho_4$ essentially does not influence the final results, and $B_w$ should be independent of $\sigma_w$ ($\sigma_w/(1+\sigma_{wL}) \approx 1$). Hence, $\sigma_w$ could be set as unchanged close to $r_{\rm acc}$. Substituting $r = r_{\rm acc}$, $\Gamma_w = \Gamma_{\rm w,acc}$ and the corresponding $\beta = \beta_{\rm w,acc}$ into Equations \ref{eq:rhowind} and \ref{eq:Bwind}, the density and magnetic field strength at the inner boundary of the blastwave can be calculated. In our calculation, $\Gamma_w = 10^3$ and $\sigma_w = 10^4$ are adopted beyond $r_{\rm acc}$. 

\begin{figure*}
\begin{tabular}{ll}
\resizebox{85mm}{!}{\includegraphics[]{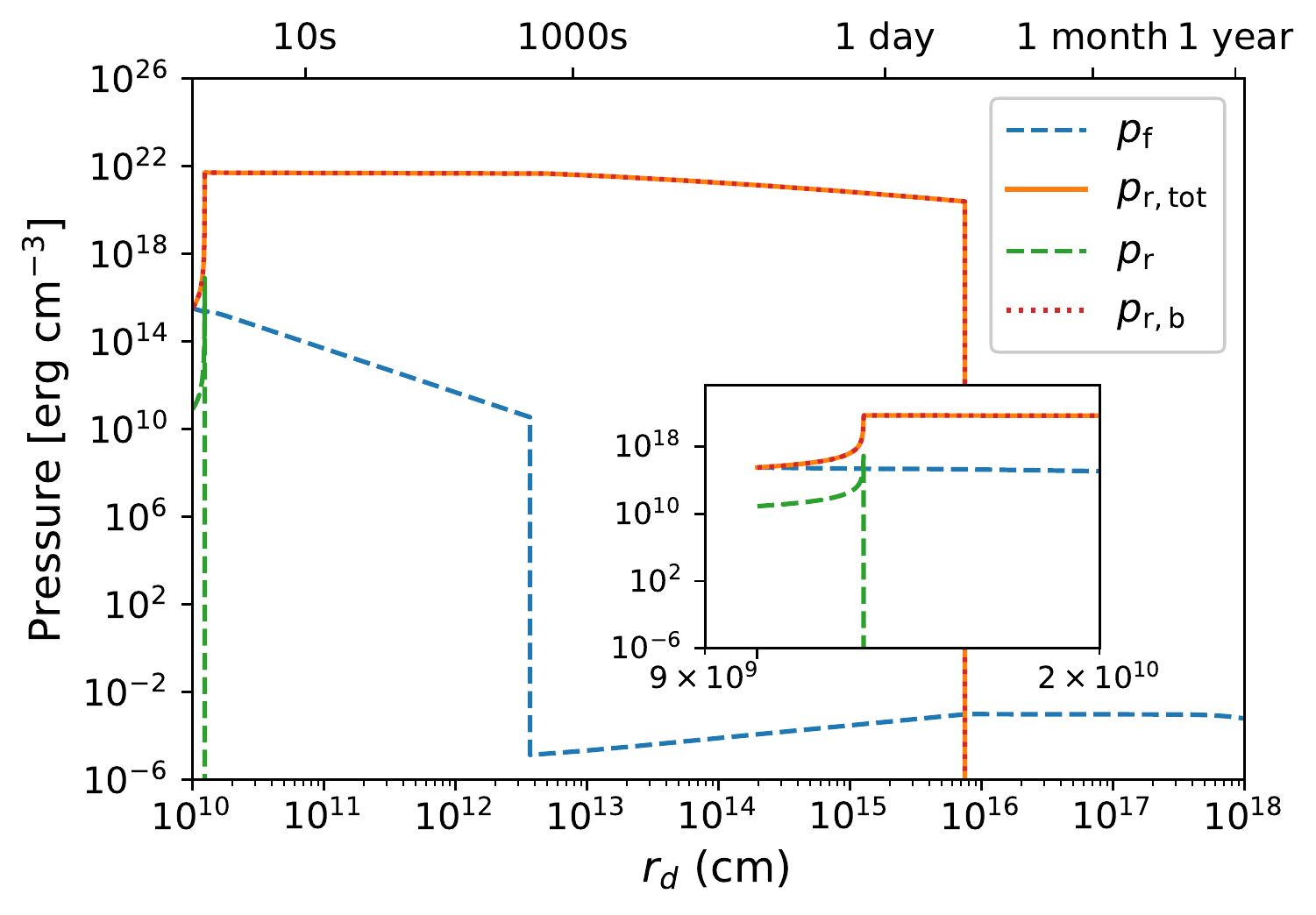}} &
\resizebox{85mm}{!}{\includegraphics[]{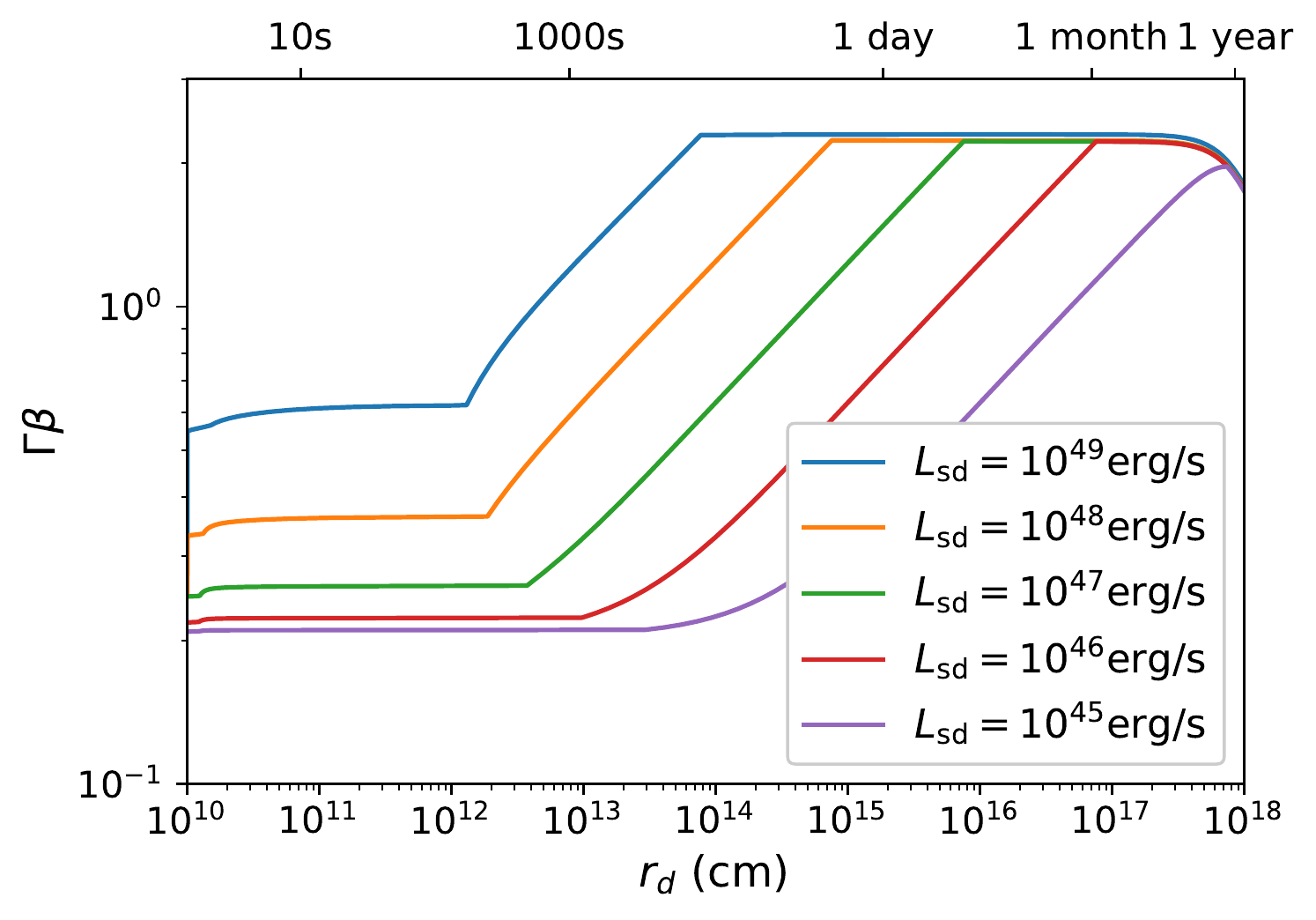}} 
\end{tabular}
\caption{Left panel: the evolution of the pressures right behind the FS and RS as a function of the contact discontinuity radius $r_d$. $L_{\rm sd} = 10^{47} {\rm erg/s}$ is adopted. The dashed lines represent the gas pressures while the dotted line represents the magnetic pressure. The total pressure is shown as the solid line. The inset is the zoom-in figure of the pressures.  Right panel: the evolution of the bulk motion four-velocity of the blastwave. $M_{\rm ej} = 10^{-3} M_{\odot}$ and $n_{\rm ISM} = 0.1 {\rm cm^{-3}}$ are adopted for both panels. The spindown luminosity $L_{\rm sd}$ is set as constant before the spindown time  $t_{\rm sd}$, which is limited by the total rotational kinetic energy of the newly born magnetar $E_{\rm rot}$ (see section \ref{sec:region4}). After $t_{\rm sd}$, the energy injection from the central engine wind becomes insignificant and we set $L_{\rm sd} = 0$. This treatment for $L_{\rm sd}$ is also valid for Figure \ref{fig:energy_fraction}, \ref{fig:pr_pg} and \ref{fig:energy}. In both panels, the upper horizontal axis shows the time ($t$) after the initial contact between the wind and the ejecta in the lab frame. The transition from Stage I to Stage II happens early at $ t < \sim 1s$ by the end of the rapid increase of the RS pressures. The transition from Stage II to Stage III happens when the forward shock pressure suddenly drops, at a time depending on the spin-down luminosity, which ranges from a few hundred seconds to a few thousand seconds. The transition times between different stages are the same in Figures \ref{fig:energy_fraction}, \ref{fig:efficiency}, \ref{fig:pr_pg} and \ref{fig:energy}.}
\label{fig:Gamma_p}
\end{figure*}

\subsection{Blastwave}
Solving the shock jump conditions and the mechanical model, the evolution of the blastwave's bulk motion velocity and the pressures right behind the FS and RS are shown in Figure \ref{fig:Gamma_p}. As shown in the figure, $p_{\rm r,tot}$ is much greater than $p_f$, so that the blastwave is undergoing acceleration because of the pressure gradient. The abrupt increase of the pressure near the RS at the beginning of the evolution (stage I) is due to the fast, inward-propagation of the RS. When the RS vanishes (stage II), the gas near the inner boundary can no longer be heated, thus $p_r = 0$. The magnetic pressure ($p_{\rm r,b}$) keeps roughly constant. During the first two stages, the bulk motion velocity for the blastwave only slightly increases, because the FS propagates slowly in the dense ejecta. After the FS breaks out of the ejecta (Stage III), $p_f$ suddenly drops and acceleration becomes significant. Later, the acceleration of the blastwave makes $p_f$ slowly increase. In stage III, analytically, the kinetic energy of the blastwave at a certain time may be estimated as
\begin{eqnarray}
E_{\rm bw,k} \approx \int_0^t 4 \pi r_r^2 p_r \beta c dt + E_{\rm bw,k,0}
\end{eqnarray}
where $E_{\rm bw,k,0}$ is the kinetic energy at the FS breakout time and the first term in the right hand side denotes the $p dV$ work. Consider that the kinetic energy (not including internal energy) of the ejecta can be generally expressed as $E_{\rm bw,k} = (\Gamma - 1)M_{\rm ej} c^2$. When $\int_0^t 4\pi r_r^2 p_r \beta c dt \ll E_{\rm bw,k,0}$, one has $E_{\rm bw,k} \sim E_{\rm bw,k,0}$ and $\Gamma$ keeps roughly a constant. When $\int_0^t 4\pi r_r^2 p_r \beta c dt \gg E_{\rm bw,k,0}$, one has 
\begin{eqnarray}
E_{\rm bw,k} &\approx& \frac{1}{8\pi}\int_{r_{d,0}}^{r_d} 4\pi r_r^2 B_r^2 dr_d \nonumber \\
&\approx& \frac{1}{8\pi}\int_{r_{d,0}}^{r_d} 4\pi r_r^2 B_4^2 dr_d \nonumber \\
&\propto& \int \Gamma^{-2} \beta^{-1} dr_d.
\end{eqnarray}
For the non-relativistic case, with $\Gamma \sim 1$ and $E_{\rm bw,k} \simeq 1/2 M_{\rm ej} \beta^2 c^2$, one gets $\beta \propto r_d^{1/2}$. For the relativistic case, with $\beta \sim 1$ and $E_{\rm bw,k} \simeq \Gamma M_{\rm ej} c^2$, one gets $\Gamma \propto r_d^{1/3}$. 

The blastwave would keep being accelerated until all the spindown energy has been injected into the blastwave. Because stage I is very short, the velocity of the RS would quickly decrease to $\beta_r \sim 0$. Therefore, this time is essentially $t_{\rm sd}$, the spindown energy of the engine. After $t_{\rm sd}$, the blastwave will experience a coasting phase. The deceleration happens roughly when the kinetic energy of the ejecta ($(\Gamma-1)M_{\rm ej}c^2$) becomes comparable with the rest mass energy of the swept ISM ($(4/3)\pi R^3 n_{\rm ISM}m_p c^2$). The deceleration radius could be then estimated as
\begin{eqnarray}
r_{\rm dec} &\approx& 1.4 \times 10^{18}~{\rm cm} \nonumber \\
&&\left(\Gamma -1\right)^{1/3} \left(\frac{M_{\rm ej}}{10^{-3}M_{\odot}}\right)^{1/3} \left(\frac{n_{\rm ISM}}{0.1~ {\rm cm^{-3}}}\right)^{-1/3},
\end{eqnarray}
which is consistent with the numerical results.

\section{Energy injection and heating efficiencies}\label{sec:efficiencies}
\subsection{Partition of the injected energy in the blastwave}
\label{sec:energy_to_blastwave}
Generally, the energy density of a magnetized fluid in the lab frame can be obtained from the 00 component of energy-momentum tensor, $T^{00} = T_{\rm FL}^{00} + T_{\rm EM}^{00}$. For fluid component, one has
\begin{eqnarray}
T_{\rm FL}^{00} = \Gamma^2 (\rho c^2 + e + p) - p = \Gamma^2 (\rho c^2 + \frac{\hat{\gamma}}{\hat{\gamma}-1}p) - p,
\end{eqnarray}
where $e = (\hat{\gamma}-1)p$ is internal energy density in the rest frame of the fluid. For the electromagnetic (EM) component, one has
\begin{eqnarray}
T_{\rm EM}^{00} = \frac{E_L^2 + B_L^2}{8\pi},
\end{eqnarray}
where $E_L$ and $B_L$ are the electric and magnetic field strength in the lab frame. Consider that in the rest frame of the magnetized fluid, there is no electric field and ${\bf B} = B {\bf e_\phi}$ in the spherical coordinates. Also, we assume the plasma is a perfect conductor that moves along $r-$ direction. The strength of electric field in the lab frame, which is induced by the motion of magnetic field, can be calculated as
\begin{eqnarray}
{\bf E_L} = -\boldsymbol{\beta} \times {\bf B_L} = - E_{L} {\bf e_{\theta}}.
\end{eqnarray}
The EM component can be then written as
\begin{eqnarray}
T_{\rm EM}^{00} = \frac{B_L^2}{4\pi} - \frac{(1 - \beta^2)B_L^2}{8\pi} = \frac{\Gamma^2 B^2}{4\pi} - \frac{B^2}{8\pi},
\end{eqnarray}
where $B = B_L/\Gamma$. The total energy density can be finally written as
\begin{eqnarray}
T^{00} &=& T^{00}_{\rm FL} + T^{00}_{\rm EM} \nonumber \\
&=& \Gamma^2 (\rho c^2 + \frac{\hat{\gamma}}{\hat{\gamma}-1} p) - p + \frac{\Gamma^2 B^2}{4\pi} - \frac{B^2}{8\pi} \nonumber \\
&=& \Gamma^2 \rho c^2 + \frac{(\Gamma^2 \hat{\gamma} - \hat{\gamma} + 1)}{\hat{\gamma}-1}p + (2\Gamma^2 - 1)\frac{B^2}{8\pi}.
\end{eqnarray}
Integrating $T^{00}$ in the entire blastwave region to find its total energy, one gets
\begin{eqnarray}
E_{\rm bw} &=& \int_{r_r}^{r_f} T^{00}dV \nonumber \\
&=& \int_{r_r}^{r_f} [\Gamma^2 \rho c^2 + (\Gamma^2 \hat{\gamma} - \hat{\gamma} + 1)\frac{p}{\hat{\gamma}-1} + (2\Gamma^2 - 1)\frac{B^2}{8\pi}]dV \nonumber \\
&\approx& \Gamma^2 \Sigma_{\rm sph} c^2 + \frac{\Gamma^2 \hat{\gamma}_{\rm eff} - \hat{\gamma}_{\rm eff} + 1}{ \hat{\gamma}_{\rm eff}-1}P_{\rm sph}  + (2\Gamma^2-1) \frac{{\cal B}_{\rm sph}}{8\pi} \nonumber 
\label{eq:E_bw}
\end{eqnarray}
The total blastwave energy ($E_{\rm bw}$) can be divided into three parts: the mass energy term ($E_{\rm bw,m} = \Gamma^2 \Sigma_{\rm sph} c^2$), the internal energy term ($E_{\rm bw,int} = \frac{\Gamma^2 \hat{\gamma}_{\rm eff} - \hat{\gamma}_{\rm eff} + 1}{ \hat{\gamma}_{\rm eff}-1}P_{\rm sph}$), and the magnetic energy term ($E_{\rm bw,b} = (2\Gamma^2-1) \frac{{\cal B}_{\rm sph}}{8\pi}$). Furthermore, the mass energy term can be divided into the rest mass energy term $E_{\rm bw,rm} = \Gamma \Sigma_{\rm sph} c^2$ and the kinetic energy term $E_{\rm bw,k} = (\Gamma - 1)\Gamma \Sigma_{\rm sph} c^2$.

At a certain time, the total energy injected into the blastwave from the inner boundary (RS or the transition radius) in the lab frame can be generally calculated as
\begin{eqnarray}
E_{\rm inj} &=& \int_0^{t} L_{\rm sd} (\beta - \beta_r) c dt \nonumber \\
&=& \int_{r_{d,0}}^{r_d} L_{\rm sd} \left(\frac{\beta - \beta_r}{\beta}\right) dr_d.
\end{eqnarray}
After the RS vanishes, one should take $\beta_r \approx 0$. Meanwhile, the shocked mass energy from Region 1 is given by
\begin{eqnarray}
E_1 &=& \int_{0}^{t} \Gamma_1^2 4\pi r_f^2 \rho_1 c^2 (\beta_f - \beta_1) c dt \nonumber \\
&=&  \int_{r_{d,0}}^{r_d} \Gamma_1^2 4\pi r_f^2 \rho_1 c^2 \left(\frac{\beta_f - \beta_1}{\beta}\right) dr_d
\end{eqnarray}
To study the energy injected from the wind only, we define the ``net'' energy of the blastwave as
\begin{eqnarray}
E_{\rm bw,net} &=& E_{\rm bw} - E_1 \nonumber \\
&=& E_{\rm bw,k,net} + E_{\rm bw,int} + E_{\rm bw,b},
\end{eqnarray}
where 
\begin{eqnarray}
E_{\rm bw,k,net} = E_{\rm bw,m} - E_1
\end{eqnarray}
is the net kinetic energy. The fraction of magnetic energy, net kinetic energy and internal energy compared to the total net energy of the blastwave are denoted as $f_b$, $f_k$ and $f_{\rm int}$, respectively.
For a correct model, the relation $E_{\rm bw,net} = E_{\rm inj}$ should always be satisfied. This is taken as an important criterion to test the validity of the mechanical model (See Appendix \ref{sec:testing}).

\begin{figure*}
\begin{tabular}{ll}
\resizebox{85mm}{!}{\includegraphics[]{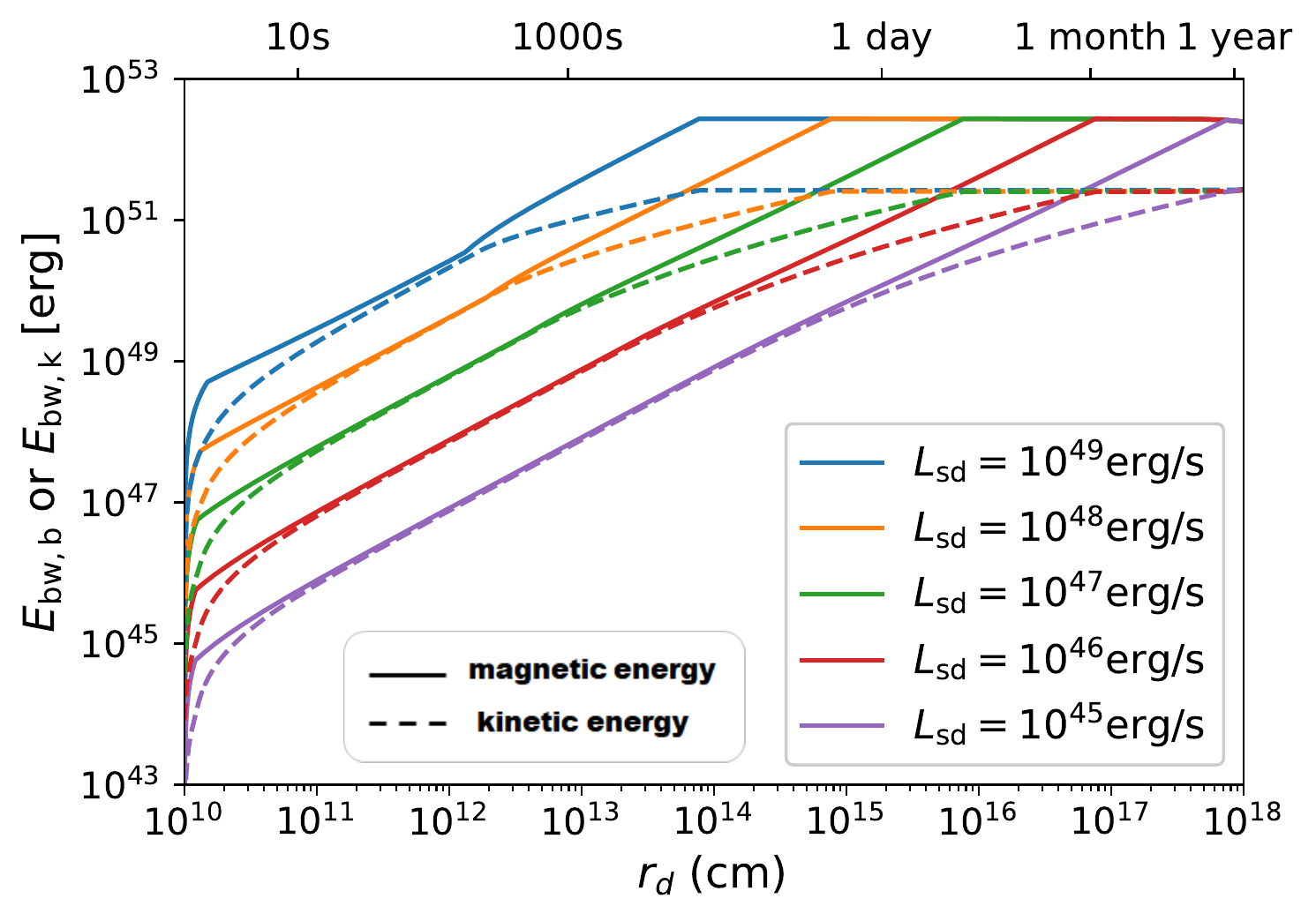}} &
\resizebox{85mm}{!}{\includegraphics[]{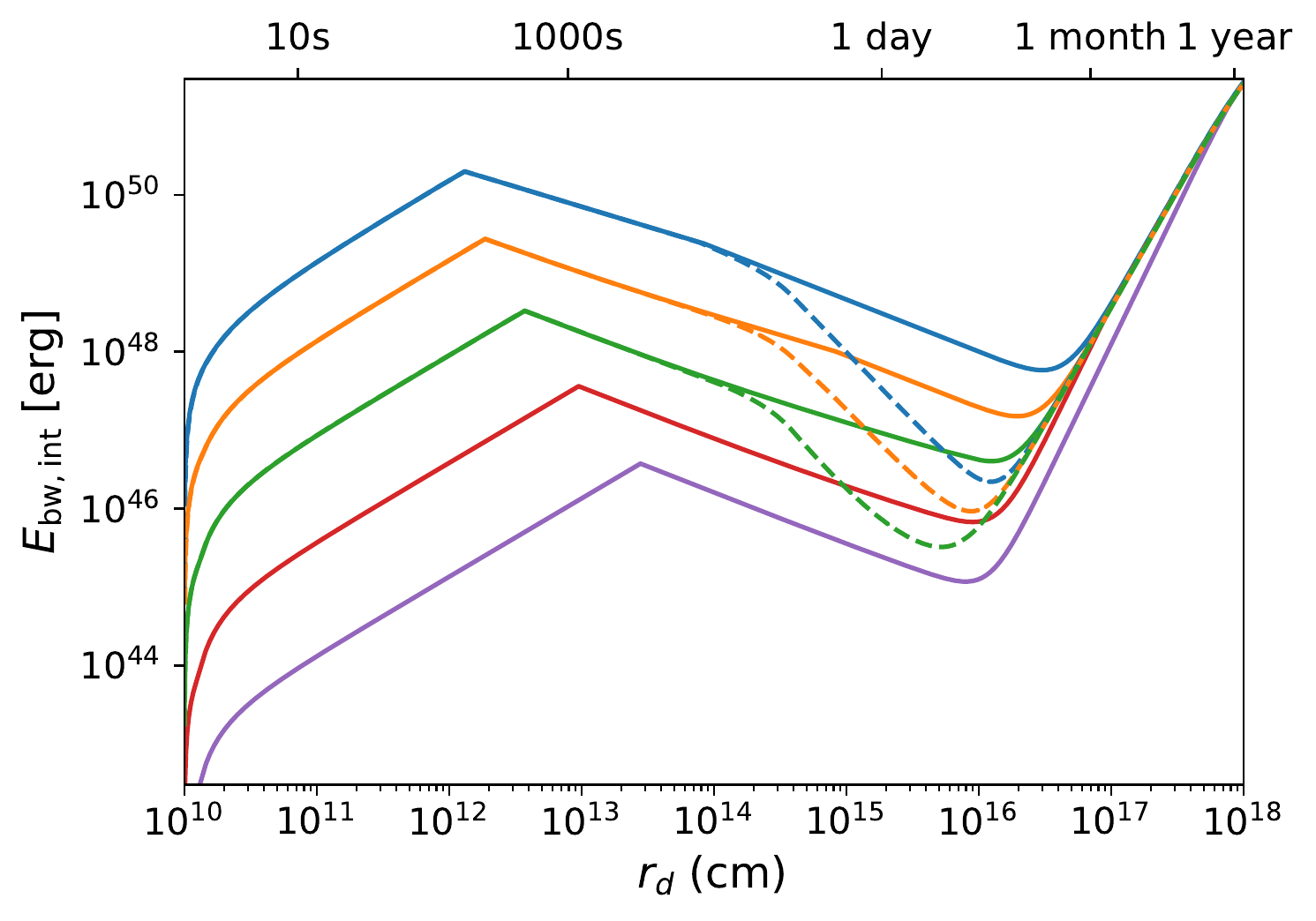}}\\
\resizebox{85mm}{!}{\includegraphics[]{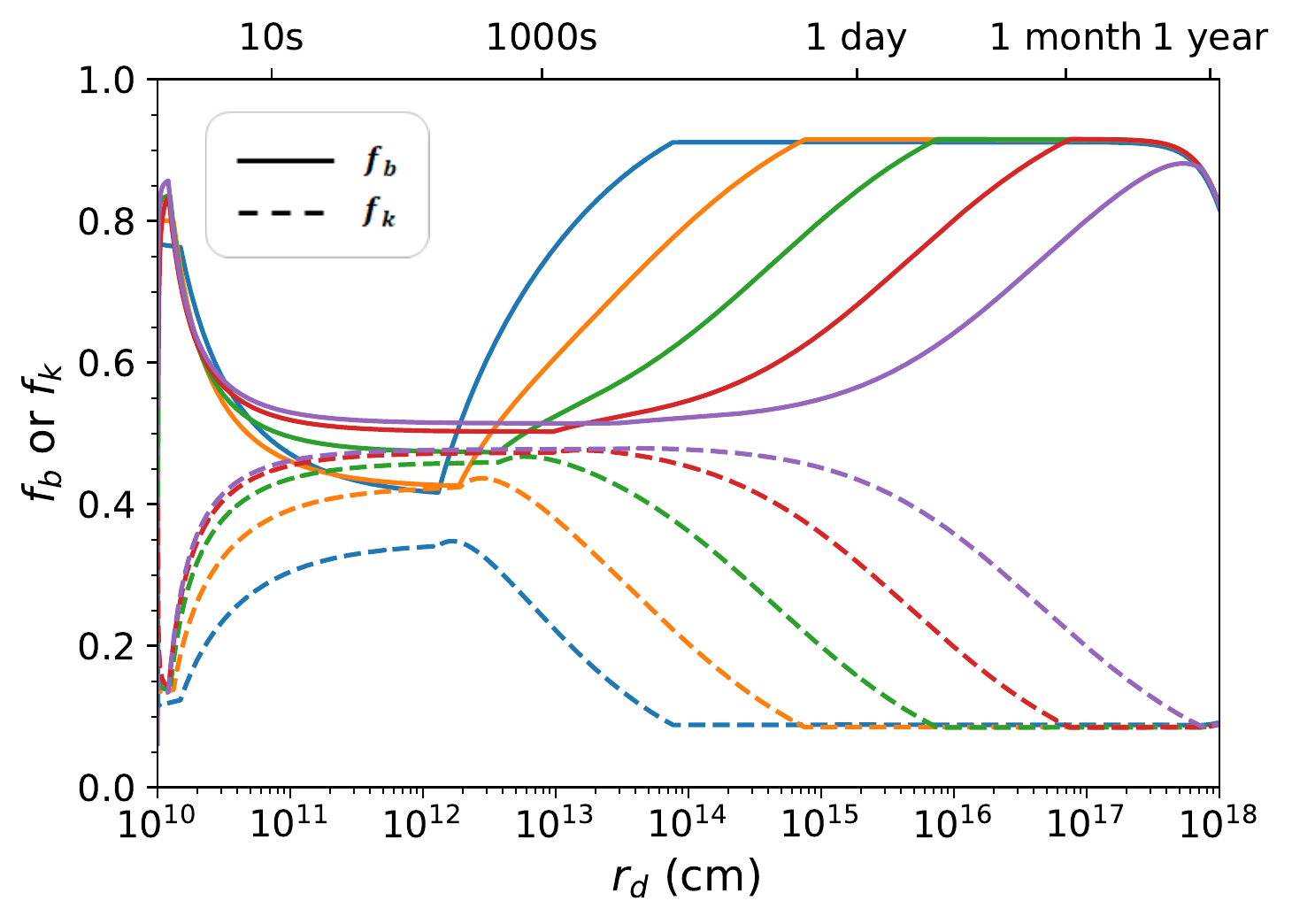}} &
\resizebox{85mm}{!}{\includegraphics[]{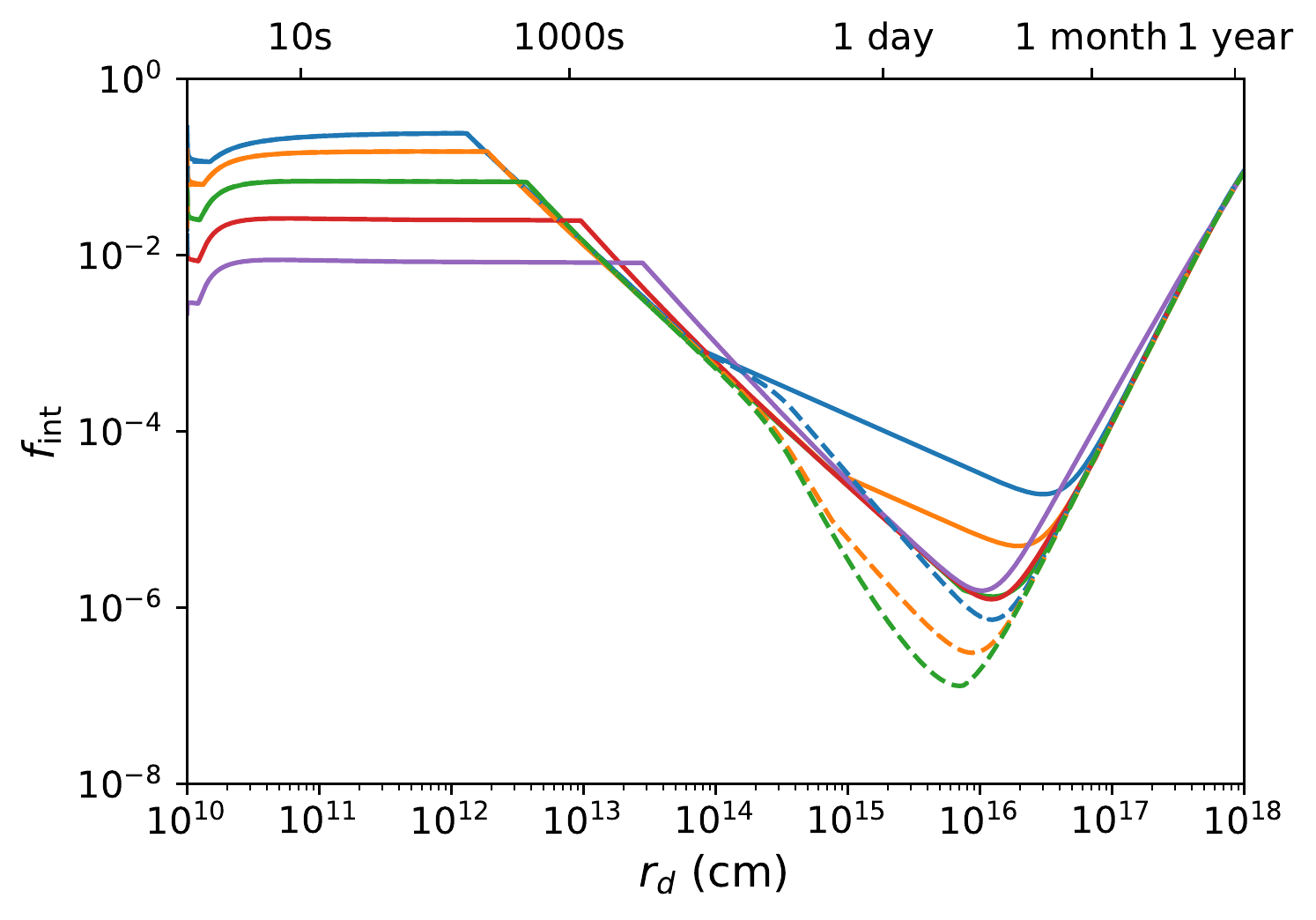}} 
\end{tabular}
\caption{The evolution of various energy components and their fractions. Upper left panel: the evolution of magnetic energy and kinetic energy of the blastwave. The solid lines represent the magnetic internal energy ($E_{\rm bw,b}$) while the dashed lines represent the kinetic energy ($E_{\rm bw,k}$). Upper right panel:  similar as the upper left panel, but the internal energy ($E_{\rm bw,int}$). Lower left panel: The evolution of the fractions of magnetic energy and kinetic energy, compared to the total energy of the blastwave. Lower right: similar to the lower left panel, but the fraction of internal energy. For the two right panels, the solid and dashed lines represent the cases with and without the radioactivity and photon emission considered, respectively. $M_{\rm ej} = 10^{-3}M_{\odot}$ is adopted for all the panels. The legends for all the panels are the same, thus are only displayed in the upper left panel.}
\label{fig:energy_fraction}
\end{figure*}

As shown in Figure \ref{fig:energy_fraction}, for our parameter sets, most of the blastwave energy is stored in the form of magnetic and kinetic energy, with the sum of the two fractions $f_b + f_k > 0.7$. Only a small fraction of energy ($f_{\rm int} < 0.3$) is stored as internal energy. In stage I, since there is a fast inward-propagating RS from which a large amount of magnetic energy is accumulated, the fractions of different energy components keep relatively stable with $f_b \sim 0.8$, $f_k \sim 0.1$, and $f_{\rm int} < 0.1$, respectively. In stage II, the RS vanishes and the accumulation of magnetic energy slows down. A good fraction of magnetic energy is converted to kinetic energy via the $p dV$ work of the magnetic pressure, thus $f_b$ decreases while $f_k$ increases. The two fractions $f_b$ and $f_k$ can be comparable. The value of $f_{\rm int}$ keeps being small and stable. In stage III, the FS breaks out of the ejecta and starts propagating in the low-density ISM. The sudden drop off of $p_f$ makes the pressure gradient near the RS and FS become larger, thus the ejecta would be accelerated significantly. However, with the acceleration of the ejecta, the efficiency for converting magnetic energy to kinetic energy would decrease ($df_K/dr_d \propto dE_{\rm bw,k}/dr_d \propto d\Gamma /dr_d \propto r_d^{-2/3}$), so that $f_k$ decreases while $f_b$ increases. Also, the rate to accumulate internal energy from the FS suddenly drops to an extremely low level. The values of $E_{\rm bw,int}$ and $f_{\rm int}$ then decrease due to the expansion of Region 2 ($dP_{\rm sph,fs}^{\prime} <  |dP_{\rm sph,exp}^{\prime}|$). Since $|dP_{\rm sph,exp}^{\prime}|$ is decreasing with $r_d$ while $dP_{\rm sph,fs}^{\prime}$ is increasing, the internal energy will rise again later when $ dP_{\rm sph,fs}^{\prime} >  |dP_{\rm sph,exp}^{\prime}|$.

The value of $f_b$ can again rise to $\sim 0.9$ at $t_{\rm sd}$, while $f_k$ decreases back to $\sim 0.1$. Then, the magnetic pressure can rarefy from the inner boundary, but would not be converted to the kinetic or internal energy of the ejecta\footnote{The evolution of magnetic energy in the blastwave after the central NS spins down is hard to calculate. Hence, we just keep magnetic energy as a constant and plot it in Figure \ref{fig:energy_fraction}.}. The kinetic energy then keeps unchanged until the ejecta reaches the deceleration radius.

\subsection{Ejecta energy-injection and heating efficiencies}
\label{sec:xi}
Two important efficiency factors that are helpful to calculate engine-fed kilonova (mergernova) emission are the total energy-injection efficiency ($\xi$) that is used to accelerate and heat up the ejecta, and the heating efficiency ($\xi_t$) in particular. Our mechanical model allows us to calculate them from the first principle.

Before the FS breaks out the ejecta (stage I and II), the energy of the ejecta includes the mass and internal energy of the blastwave and the mass energy of the unshocked ejecta ($E_{\rm usej,m}$), which is given by 
\begin{eqnarray}
E_{\rm ej} &=& E_{\rm bw,m} + E_{\rm bw,int} + E_{\rm usej,m} \nonumber \\
&=& E_{\rm bw,k,net} + E_1 + E_{\rm bw,int} + E_{\rm usej,m} \nonumber \\
&=& E_{\rm bw,k, net} + E_{\rm bw,int} + E_{\rm ej,0}
\end{eqnarray}
We can also define the ``net'' energy of the ejecta as 
\begin{eqnarray}
E_{\rm ej,net} &=& E_{\rm ej} - E_{\rm ej,0} \nonumber \\
&=& E_{\rm bw,k, net} + E_{\rm bw,int},
\end{eqnarray}
which represents the total energy injected into the ejecta from the central engine wind. Moreover, one may write
\begin{eqnarray}
dE_{\rm ej,net} &=& dE_{\rm bw,k,net} + dE_{\rm bw,int} \nonumber \\
&=& \xi L_{\rm sd} dt
\end{eqnarray}
where the parameterization factor $\xi$ is defined as the instantaneous energy injecting efficiency from the wind to the ejecta, which can be calculated as
\begin{eqnarray}
\xi &=& \frac{dE_{\rm bw,net}}{L_{\rm sd} dt}\nonumber \\ 
&=& \frac{dE_{\rm bw,k,net} + dE_{\rm bw,int}}{L_{\rm sd} dt}.
\end{eqnarray}
Similarly, the effective efficiency to heat the ejecta can be calculated as
\begin{eqnarray}
\xi_{\rm t, eff} = \frac{dE_{\rm bw,int}}{L_{\rm sd} dt}.
\end{eqnarray}
However, this is not the real heating efficiency, because the acceleration and expansion effects for the ejecta have been included in the change of $E_{\rm bw,int}$. To find the real heating efficiency ($\xi_t$), these effects have to be removed. Define $\xi_t$ as the heating efficiency in the rest frame of the ejecta, Generally, the evolution of the internal energy could be expressed as
\begin{eqnarray}
\frac{dE_{\rm bw,int}^{\prime}}{dt^{\prime}} &=& (\xi_t L_{\rm sd}^{\prime} - p^{\prime}\frac{dV^{\prime}}{dt^{\prime}})
\end{eqnarray}
where $L_{\rm sd}^{\prime} =  L_{\rm sd}$ is the spin-down luminosity in the rest frame of the ejecta. Considering
\begin{eqnarray}
\frac{dE_{\rm bw,int}^{\prime}}{dt^{\prime}} = \frac{1}{\hat{\gamma}-1} \frac{dP_{\rm sph}^{\prime}}{dt^{\prime}} = \frac{\Gamma \beta c}{\hat{\gamma}-1} \frac{dP_{\rm sph}^{\prime}}{dr_d}
\end{eqnarray}
and 
\begin{eqnarray}
p^{\prime}\frac{dV^{\prime}}{dt^{\prime}} = -\frac{P_{\rm sph}}{V} \frac{dV^{\prime}}{dr} \Gamma \beta c = - \frac{\Gamma \beta c}{\hat{\gamma}-1} \frac{dP_{\rm sph}^{\prime}}{dr_d}\bigg|_{\rm exp},
\end{eqnarray}
together with Equations \ref{eq:dP_sph} and \ref{eq:dP_fs}, one gets
\begin{eqnarray}
\xi_t L_{\rm sd}^{\prime} = \frac{\Gamma \beta c}{\hat{\gamma}-1} \left(\frac{dP_{\rm sph}^{\prime}}{dr_d}\bigg|_{\rm fs} + \frac{dP_{\rm sph}^{\prime}}{dr_d}\bigg|_{\rm rs}\right) \approx \frac{\Gamma \beta c}{\hat{\gamma}-1}\frac{dP_{\rm sph}^{\prime}}{dr_d}\bigg|_{\rm fs}
\end{eqnarray}
and 
\begin{eqnarray}
\xi_t =  \frac{4\pi r_f^2 \Gamma (\beta_f/\beta -1)p_f \beta c}{ L_{\rm sd}(\hat{\gamma}-1)}.
\label{eq:xi_t}
\end{eqnarray}
Again, in the above treatment, we have ignored the radioactive power of the neutron-rich materials and consider the heating effect only from the FS. This can allow us to understand the energy injection effect better. The radioactive heating effect will be discussed in Section \ref{sec:thermal_radiation} and considered when the lightcurves are modeled in Paper II. 

After the FS breaks out from the ejecta (stage III), the total energy of the ejecta (Region 2-2) can be expressed as 
\begin{eqnarray}
E_{\rm ej} &=& \Gamma M_{\rm ej}^{\prime} c^2+ E_{\rm ej,int} \nonumber \\
&=& \Gamma M_{\rm ej}^{\prime} c^2 + \Gamma_{\rm eff} E_{\rm ej,int}^{\prime},
\label{eq:E_ej_int}
\end{eqnarray}
where 
\begin{eqnarray}
\Gamma_{\rm eff} = \Gamma \hat{\gamma} - \frac{\hat{\gamma}-1}{\Gamma}.
\end{eqnarray}
In this stage, the mass energy accumulated from the FS would not contribute to the energy of the ejecta. The evolution of the ejecta energy should be
\begin{eqnarray}
dE_{\rm ej,net} &=& dE_{\rm ej} \nonumber \\
&=& M_{\rm ej}^{\prime} c^2 d\Gamma + E_{\rm ej,int}^{\prime} d\Gamma_{\rm eff} + \Gamma_{\rm eff} dE_{\rm ej,int}^{\prime} \nonumber \\
&=& \xi L_{\rm sd}dt,
\end{eqnarray}
because
\begin{eqnarray}
d\Gamma_{\rm eff} = \hat{\gamma} \Gamma d\Gamma + \frac{\hat{\gamma}-1}{\Gamma^2}d\Gamma
\end{eqnarray}
and
\begin{eqnarray}
dE_{\rm ej,int}^{\prime} &=& -(\hat{\gamma}-1)E_{\rm ej,int}^{\prime} \frac{dV^{\prime}}{V^{\prime}} \nonumber \\
&=& -(\hat{\gamma}-1) E_{\rm ej,int}^{\prime} (\frac{2}{r_d}dr + \frac{d\Gamma}{\Gamma}). 
\label{eq:dE_ej_int}
\end{eqnarray}
The energy injecting efficiency can be finally derived as
\begin{eqnarray}
\xi &=& \frac{M_{\rm ej}^{\prime} c^2}{L_{\rm sd}} \frac{d\Gamma}{dt} + \frac{E_{\rm ej,int}^{\prime}}{L_{\rm sd}}(\hat{\gamma}\Gamma + \frac{\hat{\gamma}-1}{\Gamma^2})\frac{d\Gamma}{dt} \nonumber \\
&-& \frac{\Gamma_{\rm eff}}{L_{\rm sd}}(\hat{\gamma}-1)E_{\rm ej,int}^{\prime} (\frac{2}{r_d}\frac{dr_d}{dt} + \frac{1}{\Gamma}\frac{d\Gamma}{dt}),
\end{eqnarray}
and the effective heating efficiency is
\begin{eqnarray}
\xi_{\rm t,eff} &=& \frac{E_{\rm ej,int}^{\prime}}{L_{\rm sd}}(\hat{\gamma}\Gamma + \frac{\hat{\gamma}-1}{\Gamma^2})\frac{d\Gamma}{dt} \nonumber \\
&-& \frac{\Gamma_{\rm eff}}{L_{\rm sd}}(\hat{\gamma}-1)E_{\rm ej,int}^{\prime} (\frac{2}{r_d}\frac{dr_d}{dt} + \frac{1}{\Gamma}\frac{d\Gamma}{dt}).
\end{eqnarray}
The real heating efficiency in this stage should be $\xi_t = 0$, because there is no internal energy being accumulated from the shocks into the ejecta.

\begin{figure}
\resizebox{80mm}{!}{\includegraphics[]{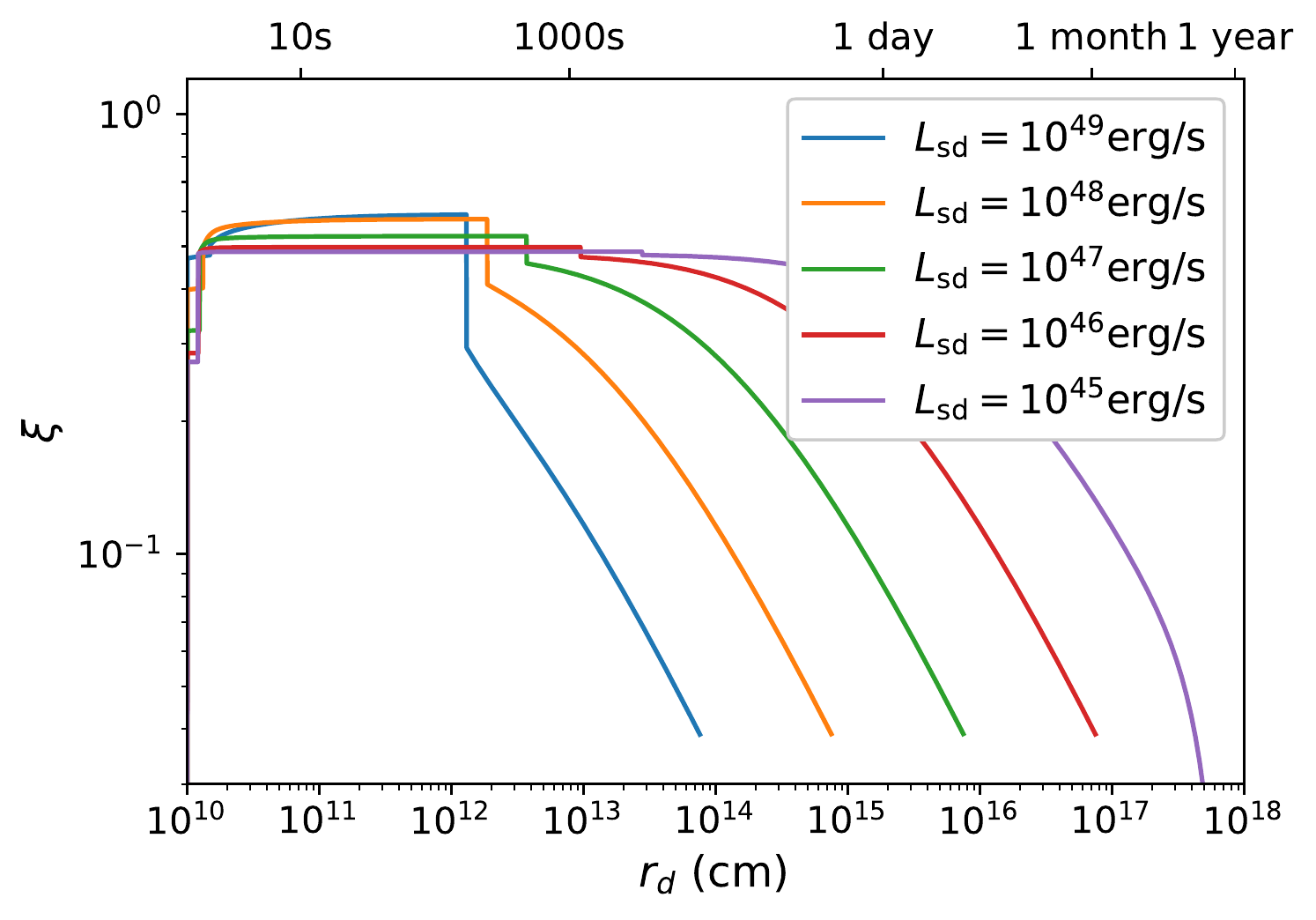}} \\
\resizebox{80mm}{!}{\includegraphics[]{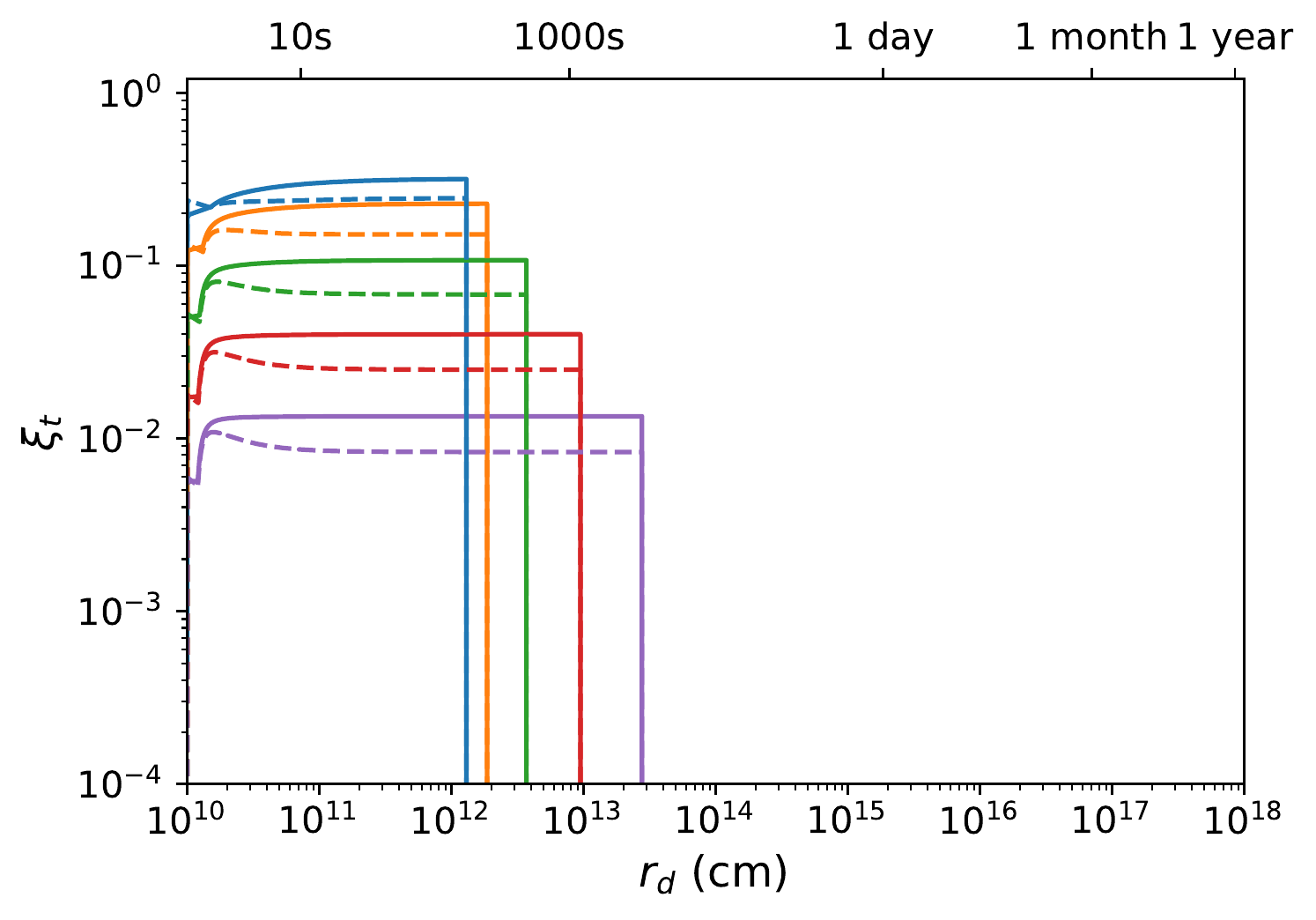}}
\caption{The evolution of the energy injecting efficiency $\xi$ (upper panel) and heating efficiency $\xi_{\rm t}$ ($\xi_{\rm t,eff}$) (lower panel). In the lower panel, solid and dashed lines represent $\xi_t$ and $\xi_{\rm t,eff}$ respectively. For both panels, $M_{\rm ej} = 10^{-3}M_{\odot}$ is adopted.}
\label{fig:efficiency}
\end{figure}

The evolution of the energy-injection and heating efficiencies are shown in Figure \ref{fig:efficiency}. In the beginning (Stage I), the total energy injection efficiency ($\xi$) is $\sim 0.3$ for all the parameter sets we discussed. Later, when the RS vanishes (Stage II), $\xi$ increases quickly to $\sim 0.5 - 0.6$. This is because $\beta_r$ changes from approximately $-1$ to $0$, thus the energy injection rate from the inner boundary decreases by half. Since the energy injected from the inner boundary is mainly magnetic energy, the energy-injection efficiency to the ejecta ($\xi$) should double.  After the FS breaks out (Stage III), $\xi$ starts to decrease all the way until the spindown time, after which the energy injection efficiency loses its meaning. The decreasing trend is consistent with the result that $f_K$ decreases with $r_d$ in Section \ref{sec:energy_to_blastwave}. The final $\xi$ values for different parameter sets are the same, i.e. $\xi \sim 0.04$, except the case that the ejecta reaches the deceleration radius before the central engine spins down (See the case with $L_{\rm sd} = 10^{45} {\rm erg/s}$). 

As shown in the lower panel of Figure \ref{fig:efficiency}, in our parameter sets, $\xi_{\rm t}$ has values ranging from $0.006$ to $0.3$ in the early stages (Stage I and II). $\xi_t$ should be slightly higher than $\xi_{\rm t,eff}$, because $\xi_{\rm t,eff}$ also includes the expansion effect. The increase at the very beginning has the same reason as that for $\xi$. Later, $\xi_t$ ($\xi_{\rm t,eff}$) suddenly decreases to $\sim 0$ (negative) at the FS breakout time (Stage III), when the FS stops heating up the ejecta and $E_{\rm ej,int}$ starts to decrease due to expansion. Since the optical depth of the shocked ISM is $\ll 1$, the random velocities for the particles in Region 2-1 cannot follow a Maxwell–Boltzmann distribution, thus would not lead to thermal emission. The internal energy of the shocked ISM should be released through synchrotron radiation, which will power a late-time broadband afterglow. In other words, the heating effect from the central engine wind vanishes after the FS breaks out the ejecta.

\subsection{effects of radioactive heating and thermal emission cooling}
\label{sec:thermal_radiation}
The radioactive power from r-process nucleosynthesis serves as a heating source for typical kilonovae. Its luminosity in the rest frame can be described with an empirical formula, which reads \citep{korobkin12}
\begin{eqnarray}
L_{\rm ra}^{\prime} = 4\times 10^{49} \left(\frac{M_{\rm ej}}{10^{-2} M_{\odot}}\right) \times \left[\frac{1}{2}-\frac{1}{\pi} {\rm arctan}\left( \frac{t^{\prime} - t_0^{\prime}}{t_{\sigma}^{\prime}} \right) \right]^{1.3} {\rm erg~s^{-1}}
\end{eqnarray}
with $t_0^{\prime} \sim 1.3~{\rm s}$ and $t_{\sigma}^{\prime} = 0.11~{\rm s}$. where $t^{\prime}$ is the time in the rest frame and $dt^{\prime} = dt/\Gamma = dr_d/\Gamma \beta c$. 

The thermal radiation of the ejecta is directly related to the kilonova/mergernova emission, through which the internal energy could be released. Therefore, it is also a cooling effect that may influence the dynamics of the blastwave. In the rest frame of the ejecta, the luminosity of thermal emission may be estimated as \citep{kasen10,kotera13}
\begin{eqnarray}
L'_e=\left \{ \begin{array}{ll} 
\Gamma E'_{\rm ej,int}c/(\tau R), & ~~\textrm{for $ \tau>1$},\\
\\
\Gamma E'_{\rm ej,int}c/R, & ~~\textrm{for $ \tau \leq 1$}.\\
\end{array} \right.
\end{eqnarray}
where $\tau = \kappa (M_{\rm ej} /V^{\prime})\Delta_{\rm ej}^{\prime}$ is the optical depth of the ejecta. The mean opacity is adopted as $\kappa = 1 \ {\rm cm^2~g^{-1}}$.

In our previous calculations, these two effects have been ignored. Here we include them in our model to test their influence on the results. Additional terms should be added to the right hand side of Equation \ref{eq:dP_sph}\footnote{In principle, source and sink terms should also be added to Equations \ref{eq:momentum_int} and \ref{eq:energy_int}. However, according to the energy partition discussed in Section \ref{sec:energy_to_blastwave} and the testing problem below in this section, for a magnetar powered kilonova (mergernova), the slight change of the internal energy is negligible compared to the total energy of the blastwave. Thus, the source and sink terms only have a very small influence on the final results.}. The term related to radioactive heating can be expressed as
\begin{eqnarray}
\frac{dP_{\rm sph}^{\prime}}{dr_d}\bigg|_{\rm ra} &=& \frac{\hat{\gamma}-1}{\Gamma \beta c} \frac{dE_{\rm ej,int}^{\prime}}{dt^{\prime}}\bigg|_{\rm ra} \nonumber \\
&=& \frac{\hat{\gamma}-1}{\Gamma \beta c} L_{\rm ra}^{\prime},
\end{eqnarray}
while the term related to the thermal emission cooling is written as
\begin{eqnarray}
\frac{dP_{\rm sph}^{\prime}}{dr_d}\bigg|_e &=&  \frac{\hat{\gamma}-1}{\Gamma \beta c} \frac{dE_{\rm ej,int}^{\prime}}{dt^{\prime}}\bigg|_e \nonumber \\
&=& - \frac{\hat{\gamma}-1}{\Gamma \beta c} L_e^{\prime} \nonumber \\
&=& - \frac{\Gamma P_{\rm sph }}{{\rm max}(\tau,1) R \beta}.
\end{eqnarray}

The results including both effects (radioactive heating and thermal emission cooling) are shown with dashed lines in the right panels of Figure \ref{fig:energy_fraction}. We only calculate the cases when $L_{\rm sd}$ is equal to $10^{49}{\rm erg/s}$, $10^{48}{\rm erg/s}$ and $10^{47}{\rm erg/s}$, respectively. With a low spindown luminosity, the radioactive power would become dominant, in which case the engine-fed mergernova could be described in the regime of a typical kilonova. In the magnetar dominated case, most of the energy injected into the ejecta is stored as kinetic energy. The radioactive heating and thermal emission cooling would not have significant influence on the dynamics of the ejecta, but would only slightly change the absolute value and fraction of internal energy. Therefore, $\xi$ should not change much. $\xi_{\rm t,eff}$ becomes no longer meaningful when more than one heating source is considered. Removing the contribution of radioactive power and thermal emission, the heating is still only (mostly) due to the FS, thus $\xi_t$ could still be calculated with Equation \ref{eq:xi_t}. Since the radioactive heating and thermal emission cooling would not influence the dynamics of the blastwave significantly, the accumulation rate of the internal energy from the FS should also not change much.

\section{Conclusions and discussion}\label{sec:conclusions}

In a series of two papers, we study the physical details of engine-fed kilonovae (mergernovae) in great detail. In this first paper focusing the dynamical evolution of the system, we performed a much more detailed study than previous work by investigating the interaction between a highly relativistic and magnetized central engine wind with a dense, massive ejecta moving with a non-relativistic speed. Such a problem is difficult to tackle purely numerically, so we adopted a semi-analytical mechanical model after slightly modifying that presented in \cite{ai21} by introducing volume integration rather than radius integration for the blastwave integrated quantities. The model is found accurate within 15\% to be compared with the numerical results for a test problem. With the mechanical model, we were able to quantify the energy-injection and heating efficiencies to the ejecta from the central engine, which have been parameterized in earlier works. Our results can be summatized as follows:
\begin{itemize}
    \item The dynamical evolution can be characterized in three stages (Figure \ref{fig:schematic}): Stage I: a pair of shocks propagate inside the ejecta and the central engine wind, respectively; Stage II: the RS vanishes after reaching the wind launching region where the magnetic pressure is large enough and the FS continues to propagate into the ejecta; and III: the FS breaks out from the ejecta and propagates in the surrounding medium.
    \item Throughout all three stages, the energy from the central engine is continuously injected into the blastwave (before the spindown timescale $t_{\rm sd}$). Since the central engine wind is Poynting-flux-dominated, most of the injected energy is stored in the form of magnetic energy. A significant fraction is converted to kinetic energy through the acceleration of the ejecta via the pressure gradient in the blastwave. Throughout the process, for $M_{\rm ej} = 10^{-3} M_\odot$, the sum of these two energy components is always greater than 70\%, i.e. $f_b + f_k > 0.7$. In the contrary, the fraction of internal energy is always smaller than 30\%, i.e. $f_{\rm int} < 0.3$.
    \item The value of ejecta energy injection efficiency, $\xi$, which is defined as the fraction of energy injected to the ejecta that is used to accelerate and heat up the ejecta, is about $0.3$ in the beginning (Stage I) and increase to $\sim 0.5 - 0.6$ after the RS vanishes (Stage II). After the FS breaks out the ejecta (Stage III), $\xi$ gradually decreases until the magnetar spins down or when the ejecta reaches the deceleration radius, with a minimum value as $\xi \sim 0.04$. 
    \item Heating of the ejecta mostly happens before the FS breaks out the ejecta. The heating efficiency, $\xi_t$, which is defined as the ratio between the internal energy accumulating rate from the FS and the spin-down luminosity, ranges from $\sim 0.006$ to $\sim 0.3$ as the spindown luminosity increases and remains constant before the FS breaks out of the ejecta. After that, $\xi_t$ abruptly drops to $0$.
\end{itemize}

There are several caveats in our treatment:
\begin{itemize}
\item The ejecta is assumed to be ``cold'' during the FS acceleration. Even if $r$-process radioactive reactions can heat up the ejecta, the temperature is usually not high enough to significantly modify the shock jump condition  \citep[e.g.][]{kumarpiran00,zhangmeszaros02}. When the hot upstream effect is important, the shock would be weakened and and heating efficiency would be suppressed compared with the cold upstream case.
\item After the FS breaks out from the ejecta (Stage III), the strength of the FS in the ISM might be underestimated. Consider that the bulk of the ejecta has been heated up by the FS while the ISM density is very low. A rarefaction wave (RW) would form in the hot ejecta, which propagates from the front part to the rear part. The front part of the ejecta would be more effectively accelerated, which would excite a stronger FS in the ISM than our estimated value. It is not easy to analytically solve a RW-FS system in the spherical geometry, because we cannot reduce $r$ and $t$ to one self-similar variable, such as the self-similar variable $a = x/t$ introduced in the Cartesian geometry case \citep{rezzolla13}). With the acceleration of the rear part of the ejecta by the magnetic pressure in Region 3, the Lorentz factor gradient would be erased at later times. Our calculation has skipped the intermediate stages in which the Lorentz factor is not supposed to be constant in space. In the late Stage III, a new FS would be formed when the pressures of the historically shocked wind decreases to a certain level, which is not included in our model. This feature should only have minor effects on the results, because, by definition, it can be treated as a substructure of the blastwave and should not have a comparable heating efficiency as the firstly formed FS when crossing the ejecta.
\item Besides shock interactions, the dissipation of magnetic fields could be another channel for energy injection. For example, according to the Internal-Collision-induced Magnetic Reconnection and Turbulence (ICMART) model proposed by \cite{zhangyan11}, the collision-triggered magnetic dissipation occurs at $r \sim 10^{15}{\rm cm}$, and part of the magnetic energy would be released through synchrotron radiation with peak energy in the X-ray band. Before the ejecta becomes transparent, X-ray photons would be trapped. In this case, the energy from the Poynting flux dominated magnetar wind could also be injected into the ejecta \citep{zhang13}. However, the efficiency for magnetic dissipation depends on the detailed magnetic reconnection mechanism, which is still unclear.  
\end{itemize}

\section*{Acknowledgements}
We thank Shangjia Zhang and Chao-Chin Yang for helpful discussion on MHD simulations. We thank He Gao for useful comments. This work is supported by the Top Tier Doctoral Graduate Research Assistantship (TTDGRA) at the University of Nevada, Las Vegas. 

%%%%%%%%%%%%%%%%%%%%%%%%%%%%%%%%%%%%%%%%%%%%%%%%%%
\section*{Data Availability}
No new data were generated or analysed in support of this research. The Python code developed to solve the theoretical problem is available upon reasonable requests. 



\bibliographystyle{mnras}
\bibliography{mybib} % if your bibtex file is called example.bib

%\begin{thebibliography}{99}
%\end{thebibliography}

%%%%%%%%%%%%%%%%%%%%%%%%%%%%%%%%%%%%%%%%%%%%%%%%%%

%%%%%%%%%%%%%%%%% APPENDICES %%%%%%%%%%%%%%%%%%%%%

\appendix

\section{Radiation pressure dominated ejecta}
\label{sec:dominant_pressure}
Because of radioactive heating and shock heating, the ejecta maintains radiation dominated throughout the evolution. We prove this using a self-consistency argument. 

\begin{figure}
\resizebox{85mm}{!}{\includegraphics[]{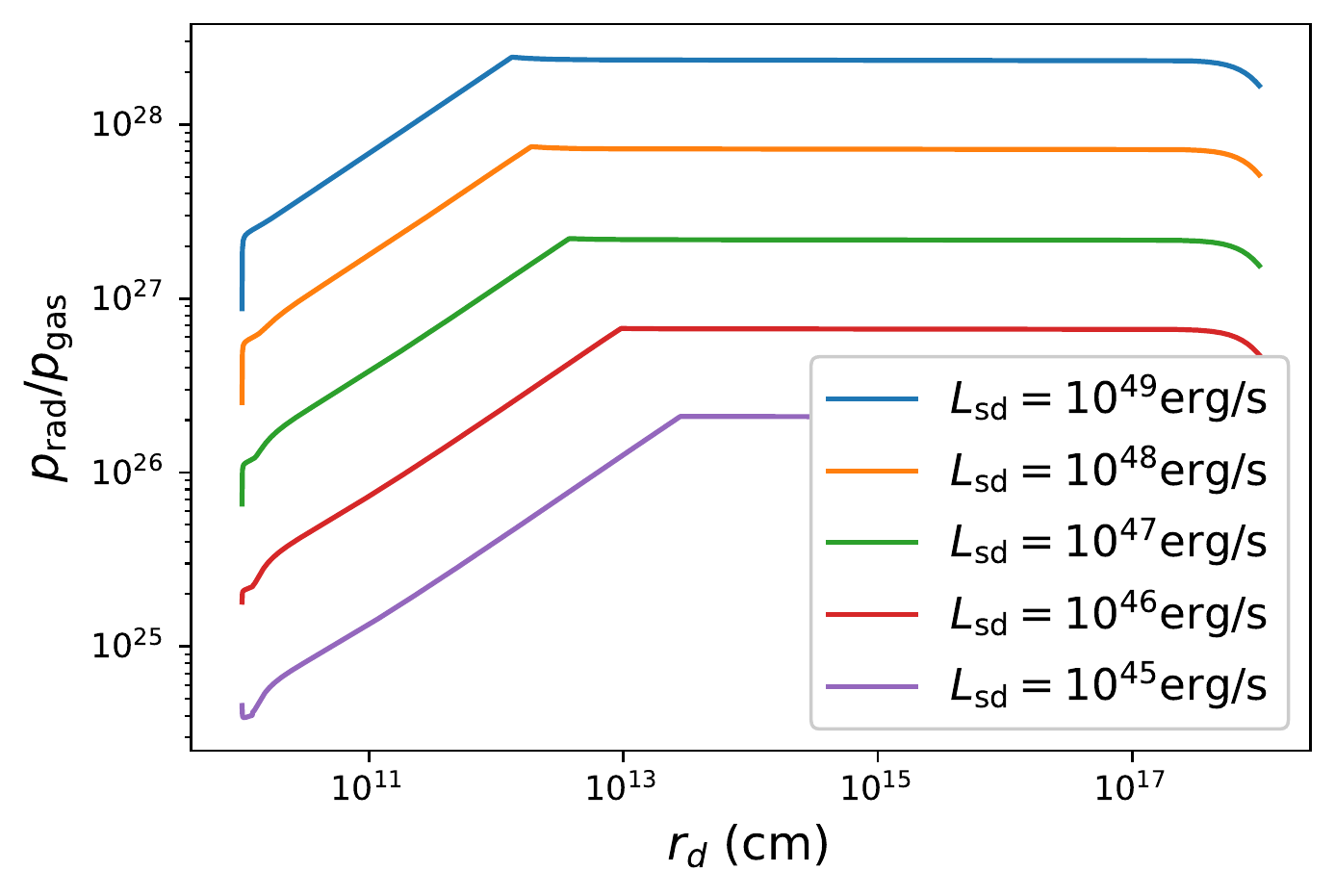}}
\caption{The ration of the mean radiation pressure and mean gas pressure in the shocked ejecta. $M_{\rm ej} = 10^3 M_{\odot}$ is adopted.}
\label{fig:pr_pg}
\end{figure}

We first assume that before the FS breakout time (Stage I and II), the radiation pressure dominates over the gas pressure in Region 2. The mean radiation pressure can be calculated as
\begin{eqnarray}
p_{\rm rad} = \frac{E_{\rm bw,int}^{\prime}}{3V_2^{\prime}},
\end{eqnarray}
where $E_{\rm bw,int}^{\prime}$ is the internal energy of the blastwave in the lab frame, as defined in Equation \ref{eq:E_bw} and the texts below the equation,
\begin{eqnarray}
V_2^{\prime} = \frac{4\pi}{3}\Gamma (r_f^3 - r_d^3)
\end{eqnarray}
is the co-moving volume of Region 2. Under the radiation pressure dominated assumption, the effective temperature $T^{\prime}$ in Region 2 should be
\begin{eqnarray}
T^{\prime} =  \left(\frac{E_{\rm bw,int}^{\prime}}{a V_2^{\prime}}\right)^{1/4},
\end{eqnarray}
where $a$ is the radiation constant. The gas pressure corresponding to this temperature can be calculated as
\begin{eqnarray}
p_{\rm gas} = n k T^{\prime} = \frac{\Gamma \Sigma_{\rm sph}}{V_2^{\prime}} k T^{\prime}
\end{eqnarray}
where $n$ stands for the mean number density in Region 2 in the rest frame, and $k$ is the Boltzmann's constant. As shown in Figure \ref{fig:pr_pg}, during this process, one has $p_r \gg p_g$.

After the FS breaks out from the ejecta (Stage III), we focus on the ejecta only (Region 2-2). The internal energy of the ejecta in its rest frame $E_{\rm ej,int}^{\prime}$ is defined in Equation \ref{eq:E_ej_int} and \ref{eq:dE_ej_int}. In this stage, again let us assume that the radiation pressure dominates over the gas pressure in the ejecta, the mean radiation pressure can be calculated as
\begin{eqnarray}
p_{\rm rad} = \frac{E_{\rm ej,int}^{\prime}}{3V_{\rm ej}^{\prime}},
\end{eqnarray}
where 
\begin{eqnarray}
V_{\rm ej}^{\prime} = \frac{4\pi}{3}\Gamma \left[(r_d + \Delta_{\rm ej,s})^3 - r_d^3\right].
\end{eqnarray}
is the co-moving volume of the ejecta. $\Delta_{\rm ej,s}$ represents the thickness of the ejecta in the lab frame at the FS breakout time, which can be recorded when numerically solving the dynamical evolution of the blastwave. Similarly, the effective temperature in the ejecta can be calculated as
\begin{eqnarray}
T^{\prime} =  \left(\frac{E_{\rm ej,int}^{\prime}}{a V_{\rm ej}^{\prime}}\right)^{1/4}.
\end{eqnarray}
The gas pressure in the ejecta which is related to this temperature should be
\begin{eqnarray}
p_{\rm gas} = n k T^{\prime} = \frac{M_{\rm ej}}{V_{\rm ej}^{\prime}} k T^{\prime}.
\end{eqnarray}
Again, as shown in Figure \ref{fig:pr_pg}, $p_r \gg p_g$ is also satisfied in this stage. Therefore, the ejecta should be radiation dominated throughout the evolution stages.

\section{Thickness of region 2}
To make Equation \ref{eq:r_p} valid, the thickness of region 2 should be much smaller than the radius of the contact discontinuity. Denote thickness of region 2 in the lab frame as $\Delta_2$, which can be expressed as
\begin{eqnarray}
\Delta_2 (t) = \int_{0}^{t}(\beta_{\rm fs}(t) - \beta (t)) dt,
\end{eqnarray}
where $t$ stands for the time in the lab frame since the formation of the FS-RS pair. We can also write down the position of contact discontinuity as
\begin{eqnarray}
r_d (t) = r_0 + \int_0^{t} \beta (t) dt 
\end{eqnarray}
Calculate their ratio we find 
\begin{eqnarray}
\frac{\Delta_2}{r_d} < \frac{\int_{0}^t (\beta_{\rm fs} - \beta)dt}{\int_{0}^{t} \beta dt} < {\rm max} \left(\frac{\beta_{\rm fs} - \beta}{\beta}\right).
\end{eqnarray}
The propagation of the shock can be written as a function of the shock strength (e.g. $u_{2s} = f(\Gamma_{21})$). Specifically, in our problem, when the upstream (region 1) is cold and motionless, $\Gamma_{21} = \Gamma $ and
\begin{eqnarray}
u_{\rm 2s}^2 = \frac{(\Gamma - 1)(\hat{\gamma}-1^2)}{\hat{\gamma}(2 - \hat{\gamma})(\Gamma-1)+2}.
\end{eqnarray}
with the adiabatic index of the downstream materials $\hat{gamma}$ ranging from $4/3$ to $5/3$. Then follow Equation \ref{eq:beta_s}, the velocity of the FS propagating in region 1 can be calculated. We find that $(\beta_{\rm fs} - \beta)/\beta < 0.33$ for $\hat{\gamma} = 5/3$ and $(\beta_{\rm fs} - \beta)/\beta < 0.16$ for $\hat{\gamma} = 4/3$. It monotonically decreases with $\Gamma$. Since region 2 is dominated by radiation pressure (see Appendix \ref{sec:dominant_pressure}), thus $\hat{\gamma} = 4/3$ can be adopted. Finally, we claim $\Delta_2/r_d < 0.16$. When region 2 is accelerated to a relativistic bulk motion ($\Gamma > 2$), $\Delta_2/r_d$ can be smaller than $0.05$. Although the thin region 2 assumption is not always perfectly valid, it is still an acceptable approximation.

\section{A test on the mechanical model with energy conservation}
\label{sec:testing}
\begin{figure*}
\begin{tabular}{ll}
\resizebox{85mm}{!}{\includegraphics[]{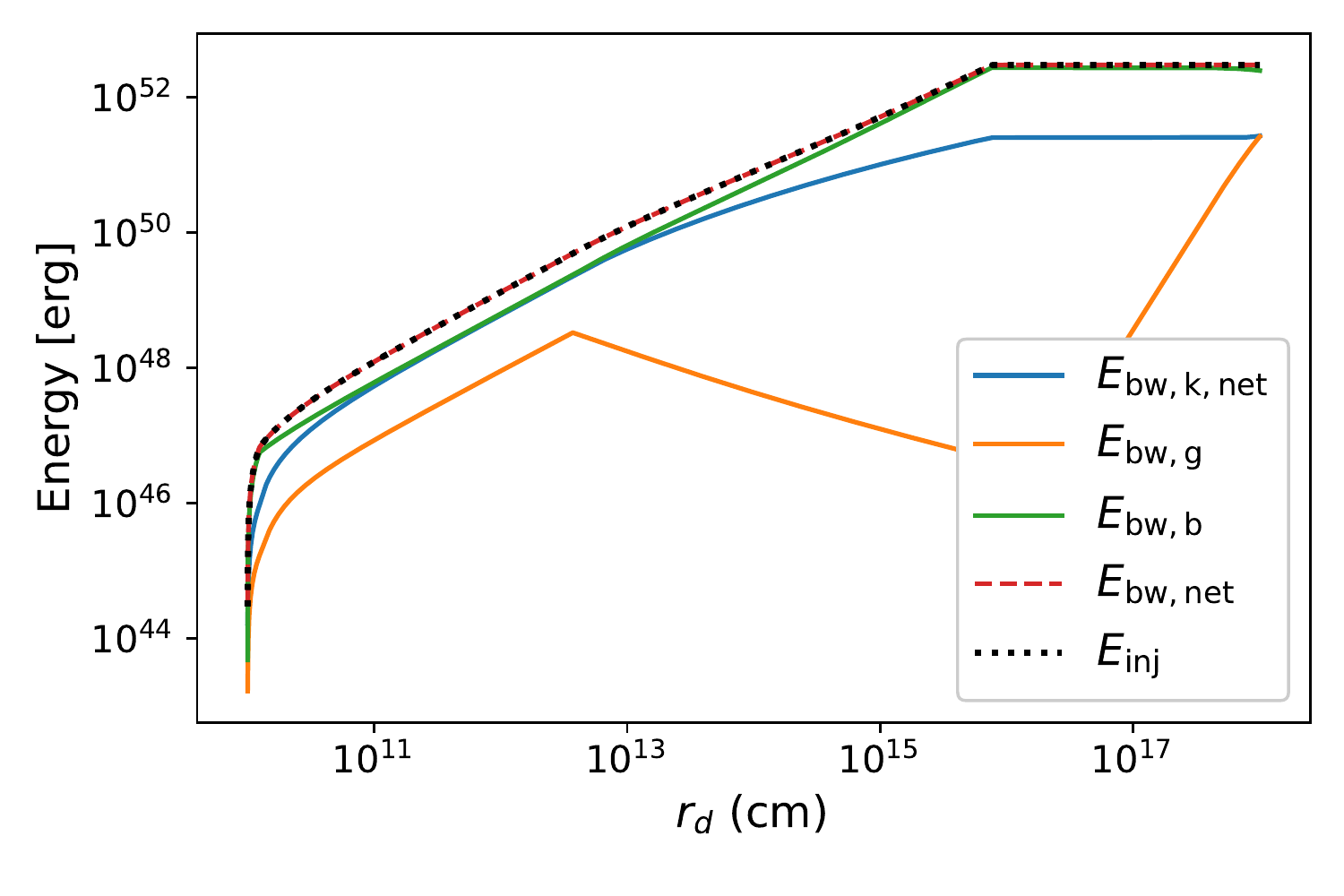}} &
\resizebox{85mm}{!}{\includegraphics[]{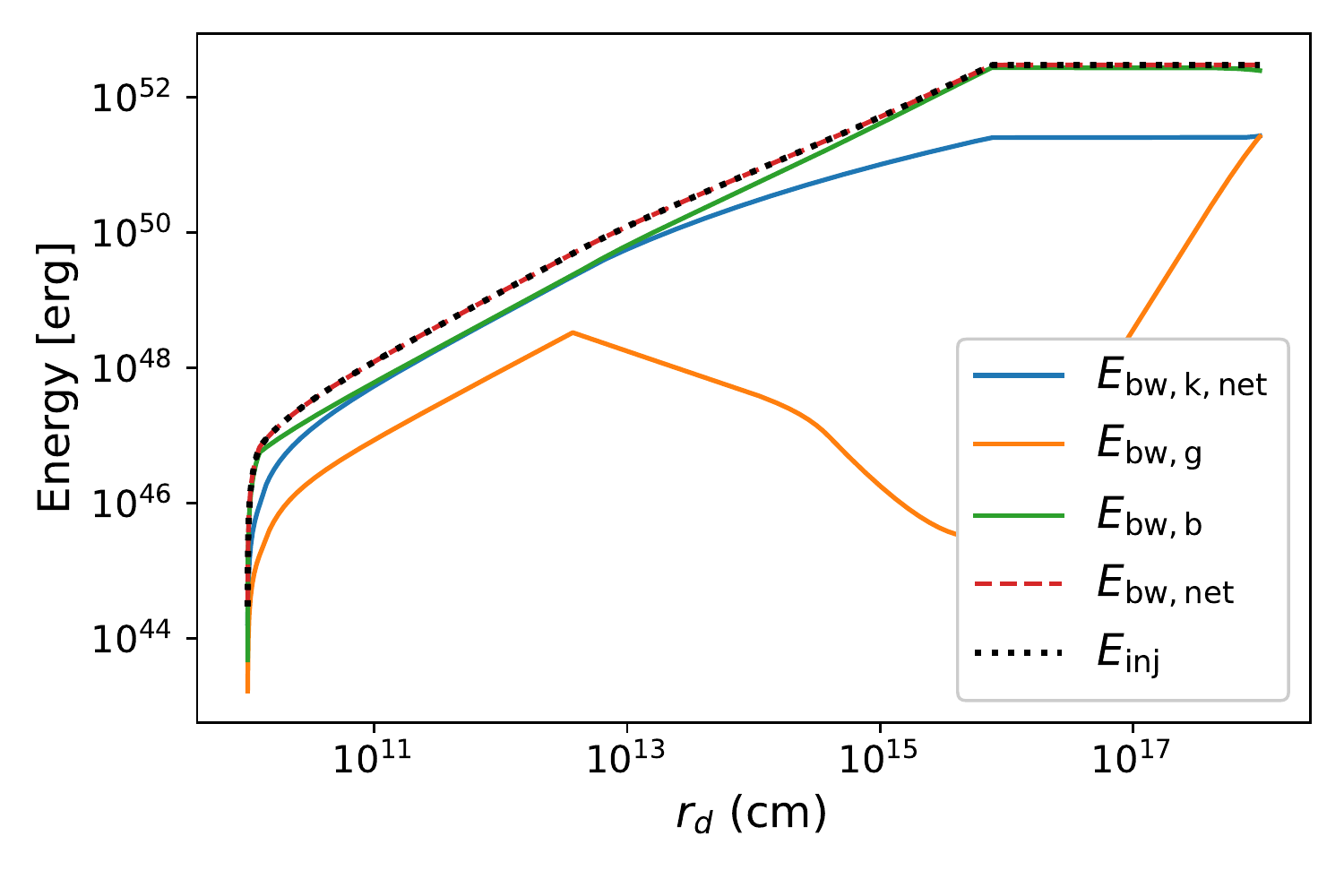}}
\end{tabular}
\caption{The evolution for different energy components and the energy injected from the inner boundary, with both radioactive power and  thermal emission considered (right panel), or without either of them considered (left panel). For both panels, $M_{\rm ej} = 10^3 M_{\odot}$ is adopted.}
\label{fig:energy}
\end{figure*}
As shown in Figure \ref{fig:energy}, the curve for $E_{\rm bw,net}$ always coincide with the curve for $E_{\rm inj}$, which means that at any time the energy conservation law is satisfied. Also, there is no significant difference on the $E_{\rm bw,net}$ curves whether or not the radioactive power and thermal emission and cooling are considered, but the difference is obvious on $E_{\rm bw,int}$ curves. This indicates that these effects could change the evolution of the internal energy, but would not influence the dynamics of the blastwave.

% Don't change these lines
\bsp	% typesetting comment
\label{lastpage}
\end{document}